\documentclass[acmsmall]{acmart}

\AtBeginDocument{%
  }

\title{Developer Perspectives on Licensing and Copyright Issues Arising from Generative AI for Software Development}

\setcopyright{cc}
\setcctype{by-nc-sa}
\acmJournal{TOSEM}
\acmYear{2025} \acmVolume{1} \acmNumber{1} \acmArticle{1} \acmMonth{1} \acmPrice{}\acmDOI{10.1145/3743133}

\author{Trevor Stalnaker}
\orcid{0009-0005-6000-4227}
\affiliation{%
  \institution{William \& Mary}
  \city{Williamsburg, VA}
  \country{USA}
}
\email{twstalnaker@wm.edu}

\author{Nathan Wintersgill}
\orcid{0009-0006-2123-7412}
\affiliation{%
  \institution{William \& Mary}
  \city{Williamsburg, VA}
  \country{USA}
}
\email{njwintersgill@wm.edu}

\author{Oscar Chaparro}
\orcid{0000-0003-2838-685X}
\affiliation{%
  \institution{William \& Mary}
  \city{Williamsburg, VA}
  \country{USA}
}
\email{oscarch@wm.edu}

\author{Laura A. Heymann}
\orcid{0000-0001-9258-2105}
\affiliation{%
  \institution{William \& Mary}
  \city{Williamsburg, VA}
  \country{USA}
}
\email{laheym@wm.edu}

\author{Massimiliano Di Penta}
\orcid{0000-0002-0340-9747}
\affiliation{%
  \institution{University of Sannio}
  \city{}
  \country{Italy}
}
\email{dipenta@unisannio.it}

\author{Daniel M German}
\orcid{0000-0001-5661-4392}
\affiliation{%
  \institution{University of Victoria}
  \city{Victoria}
  \country{Canada}
}
\email{dmg@uvic.ca}

\author{Denys Poshyvanyk}
\orcid{0000-0002-5626-7586}
\affiliation{%
  \institution{William \& Mary}
  \city{Williamsburg, VA}
  \country{USA}
}
\email{denys@cs.wm.edu}

\date{May 2025}

\usepackage{ifthen}

\newboolean{showcomments}
\setboolean{showcomments}{true}

\newboolean{showboxes}
\setboolean{showboxes}{true}

\usepackage{verbatim}

\usepackage[utf8]{inputenc}
\usepackage{url}
\usepackage{colortbl}
\usepackage{balance}
\usepackage{xspace}
\usepackage{graphicx}
\usepackage{bold-extra}
\usepackage{paralist}
\usepackage{multirow}
\usepackage{listings}
\usepackage{boxedminipage}
\usepackage{soul}
\usepackage{enumitem}
\usepackage[normalem]{ulem}
\setlist{nolistsep,leftmargin=.5cm}
\usepackage{booktabs}
\usepackage{tabularx}
\usepackage{svg}
\usepackage{calc}
\usepackage{float}
\usepackage{multirow}
\usepackage[normalem]{ulem}
\useunder{\uline}{\ul}{}
\usepackage[most]{tcolorbox}
\usepackage{caption}
\usepackage{microtype} 
\usepackage{algorithmic}
\usepackage{textcomp}
\usepackage{xcolor}

\usepackage{wrapfig}
\usepackage{subcaption}
\usepackage{cleveref} 
\usepackage{adjustbox}
\usepackage{totcount}

\usepackage{breakurl}

\usepackage{hyperref}

\ifthenelse{\boolean{showcomments}}
{\newcommand{\nb}[2]{
		\fbox{\bfseries\sffamily\scriptsize#1}
		{\sf\small$\blacktriangleright$\textit{#2}$\blacktriangleleft$}
	}
	
}
{\newcommand{\nb}[2]{}
	
}

\newcommand{\ie}{\textit{i.e.},\xspace}
\newcommand{\eg}{\textit{e.g.},\xspace}

\newcommand{\etal}{\textit{et al.}\xspace}

\newcommand{\resp}[1]{$R_{#1}$}
\newcommand{\iresp}[1]{$RI_{#1}$}

\newcounter{findingcounter}
\newtotcounter{totalfindings}
\ifthenelse{\boolean{showboxes}}
{
    \newcommand{\finding}[1]{%
      \refstepcounter{findingcounter}
      \begin{tcolorbox}[boxsep=1pt,left=2pt,right=2pt,top=1pt,bottom=1pt]%
      \textbf{Finding \arabic{findingcounter}:} #1
      \end{tcolorbox}%
      \addtocounter{totalfindings}{1}
    }
    
  }
{
    \newcommand{\finding}[1]{}
}

\begin{abstract}

Despite the utility that Generative AI (GenAI) tools provide for tasks such as writing code, the use of these tools raises important legal questions and potential risks, particularly those associated with copyright law.  
As lawmakers and regulators respond to these questions, the views of users can offer relevant perspectives. 
In this paper, we provide: (1) a survey of 574 developers on the licensing and copyright aspects of GenAI for coding, as well as follow-up interviews; (2) a snapshot of developers' views at a time when GenAI and perceptions of it were rapidly evolving; and (3) an analysis of developers' perspectives, yielding insights and recommendations that can inform future regulatory decisions in this evolving field. Our results show the benefits developers derive from GenAI, how they view the use of AI-generated code as similar to using other existing code, the varied opinions they have on who should own or be compensated for such code, that they are concerned about data leakage via GenAI, and other findings, providing organizations and policymakers with valuable insights into how the technology is being used and the concerns that stakeholders believe warrant attention.

\end{abstract}

\begin{document}




\maketitle

\footnotesize{The first two authors contributed equally to this work.}

\section{Introduction}
\label{sec:intro}
Generative AI (GenAI) tools have been widely adopted in many different domains, including software engineering (SE)
~\cite{li2023starcoder, ebert2023generative, rajbhoj2024accelerating, all_using_ai}. Several GenAI coding assistants, including GitHub Copilot~\cite{copilot}, Tabnine~\cite{tabnine}, Codeium~\cite{codeium}, and Cody~\cite{cody}, as well as general-purpose tools such as ChatGPT~\cite{chatgpt}, Claude~\cite{anthropic_models}, and Gemini~\cite{geminiteam2023gemini}, %
have become readily accessible, either as IDE extensions or stand-alone applications, enabling developers to perform many coding tasks, including automated code completion, summarization, and debugging. 
A 2024 Stack Overflow survey reported that 76 percent of 65,000 participating developers were using or planning to use AI coding tools \cite{stackoverflow_survey}; more than one million developers used Copilot \cite{copilot} in its first year~\cite{copilotAdoption}.  The popularity of these tools comes from the fact that they are easy to access and can generate various types of content with relatively little effort, thus helping users perform many tasks more efficiently~\cite{liang2024large, barke2023grounded}.

In addition to these benefits, however, the use of GenAI tools can introduce various issues, including bias or discrimination~\cite{ferrara2023fairness}, security threats~\cite{majdinasab2024assessing}, and compromise of private information~\cite{gupta2023chatgpt}. GenAI can also pose legal risks related to intellectual property concerns, including copyright infringement~\cite{congressCopyright}. Many legal questions regarding the use of GenAI tools remain unanswered, and any answers will likely vary among jurisdictions and across cases. 

At the time of writing, there are several pending lawsuits against high-profile providers of GenAI tools including OpenAI~\cite{authorsVsChatGPT23}, StabilityAI~\cite{gettyVsStability23}, and Midjourney~\cite{andersenVsStability22}, as well as active consideration by the U.S. Copyright Office and other governmental entities around the world of the copyright issues raised by the use of GenAI~\cite{UCOAIstudy,usCopyRegis23,aiact2024, copyright_office_2}. These matters involve several considerations, including the extent to which works associated with the use of GenAI, including prompts and models, are protected by copyright and, if so, who owns the rights to such works; whether the use of open-source software and other protected works as training data results in license violations or otherwise constitutes copyright infringement; and whether output resulting from the use of GenAI that is similar to preexisting work constitutes copyright infringement. 

Because any regulation of this space---whether by governmental entities or corporate entities---should take account of current and future practices, we aim in this article to identify developers' views on (1) using GenAI tools for software development tasks and (2) associated legal concerns, focusing specifically on copyright law. This work aligns with requests from the U.S. Copyright Office seeking the opinions of experts, organizations, and individuals (\eg developers) for active consideration in the development of future regulation \cite{copyright_office_NOI}.

This article presents a study conducted by a joint team of SE and legal researchers that surveyed 574 software developers worldwide who reported using GenAI tools for software development tasks (particularly code generation). Through an online survey and follow-up interviews, we investigated developers' opinions on potential emerging legal issues, the perception of what is copyrightable, ownership of generated code, and related considerations. We also sought to understand potential developer misconceptions, learn about the impact of GenAI on their work, and assess their awareness of licensing/copyright risks. 

Using qualitative and quantitative methods to analyze survey responses, we found that developer opinions on copyright issues vary broadly, particularly on the topic of model output ownership, and that many developers are aware of the nuances and complications associated with answering these complex legal questions. We discuss the results of our study, interpret the findings against the backdrop of U.S. law, and offer insights for future work. We focus on developers rather than other stakeholders, such as compliance officers and policymakers, because developers are often the ones who directly interact with GenAI during software development. We survey an international population of developers because software development is a global activity, particularly for open-source projects. In addition, large technology companies regularly employ individuals from various countries (and jurisdictions) around the world.  Given this, it is appropriate to consider the thoughts, opinions, and perspectives of developers regardless of their geographical location.

We make available our survey, results, and other artifacts for transparency and validation in an online replication package~\cite{anonymous_repo}.

\section{Background and Related Work}
\label{sec:back}

Our work is intersectional, based on concepts drawn both from the legal domain and from software engineering. We discuss these concepts in this section as well as prior work related to our study.

\subsection{Legal background}
\label{sec:legal_background}
Unless it is in the public domain, computer code is generally protected by copyright in the U.S. so long as it constitutes an original work of authorship and is fixed in a tangible medium of expression, such as on a server~\cite{usc102a}. Similar provisions exist in the intellectual property legislation of other countries, including those in the European Union \cite{eucopyrights}.  The owner of a copyright---which may be one or more individuals or corporate entities---has several exclusive rights under U.S. copyright law, including the right to reproduce the work in copies, to create derivative works (works based on the copyrighted work), and to distribute the work, which also includes the right to authorize others to engage in such activities~\cite{usc106}. In addition, U.S. copyright law prohibits the removal or alteration of ``copyright management information'' defined as including ``the name of, and other identifying information about, the author of a work,'' although the limits of this provision and its application to GenAI are as yet uncertain \cite{usc1202b}.%

The use of GenAI implicates these rights in several ways that are currently under consideration by the U.S. Copyright Office and the courts.~\cite{usCopyRegis23, 2023chatgpt, author_lawsuit, copyrightRegistration}.  For example, the plaintiffs in \textit{Doe v. GitHub} \cite{doeVsGithub22}, filed in November 2022, allege that Copilot removes copyright management information and violates the copyright licenses of the plaintiffs' open source software by using that code as training data and generating code that is a near-identical reproduction of the plaintiffs' code but without adhering to the terms of the associated licenses. \looseness=-1

The resolution of this and similar cases depends on both technological and legal questions. On the technological side, liability may depend on whether authors can successfully show that \textit{their works} are indeed included in a particular LLM's training data, as well as whether they can show that those works are reproduced or memorized in a way that would trigger liability under copyright law\cite{cooper2024files}. On the legal side, liability may depend on whether and how the generated content is similar to any particular author's work.  %
For example, generated content that is similar to previous works only with respect to the underlying concept or function is not infringing under U.S. copyright law because protection does not extend to any ``idea, procedure, process, system, method of operation, concept, principle, or discovery''~\cite{usc102b}. Additionally, some legal scholars argue that the use of a %
work protected by copyright as part of a model's training data may be deemed a fair use under U.S. law and so not infringing \cite{lemley2020fair}, similar to how Google's use of copyrighted material in the database for its Google Books search engine was held to be fair use by the U.S. Court of Appeals for the Second Circuit in 2015~\cite{2015authors}. %

Legislation both within and outside the U.S. also has implications for GenAI and copyright. Various U.S. states have attempted to enact legislation to govern GenAI \cite{colorado_bill, california_regulation}. In March 2024, the European Parliament approved the EU AI Act~\cite{aiact2024}, which, among other things, requires the providers of general-purpose AI models to ``put in place a policy to comply with Union law on copyright and related rights, and in particular to identify and comply with, including through state-of-the-art technologies, a reservation of rights expressed pursuant to Article 4(3) of Directive (EU) 2019/790'' \cite{aiact_article53}. This requires that creators of general-purpose AI models respect upstream copyrights---use due diligence in using copyrighted materials as part of their training data, obtain consent from copyright owners, and allow copyright holders to opt out---and respect downstream copyrights, mitigating the risk---often due to overfitting---of generating verbatim content that is too similar to the (copyrighted) original. In 2023, an early announcement of the EU Global Principles for Artificial Intelligence, which led to the EU AI Act, encouraged some organizations, including French media organizations France Médias Monde and TF1, to forbid companies such as OpenAI from mining their data \cite{FrenchCase2023}; that year also saw a temporary ban on ChatGPT in Italy~\cite{ChatGPTItalyBan}. %
Even countries like Japan, which initially affirmed that training AI models on protected content was not a violation of copyright law \cite{japanAI}, have started to consider additional restrictions designed to protect copyright holders \cite{JapanNewsAIPolicy23}.  %
Gervais \etal \cite{gervais2024heart} provide an in-depth overview of the then-current copyright landscape around the globe.

\looseness=-1

\subsection{Technical background}
\label{sec:technical_background}
Prior work relevant to GenAI and copyright issues focuses on research that may help answer some of the technological questions raised above. For example, there are techniques to potentially reveal whether a machine learning model was trained on a certain data record, such as membership inference attacks~\cite{duan2024membership, hu2022membership, majdinasab2024trained, zhang2023code, yang2023gotcha}. Such attacks can extract training data from the output of the target model, including private and personal information~\cite{carlini2021extracting}. Memorization of training data is a known problem, including in models trained on code~\cite{yang2024unveiling}, with research suggesting that typical measures aimed at preventing verbatim extraction of training data are not sufficient to prevent the data from being obtained with other methods, such as by using style-transfer prompts~\cite{ippolito2023preventing}. %
Tools such as CopyrightCatcher \cite{patronusCopyrightCatcher} have been developed to detect the presence of copyrighted material in the output of models, demonstrating that this is an issue of concern for developers.

\subsection{Related work}
Prior %
research has explored several questions and concerns related to copyright and GenAI~\cite{lucchi2023chatgpt, guadamuz2024scanner}. For example, Lee \etal~\cite{lee2023talkin} outline the intricate GenAI supply chain and its intersection at various points with aspects of U.S. copyright law, revealing that answers to copyright questions may depend on the specific details of how an individual model was trained. Craig \cite{craig2024ai} has cautioned against applying or expanding copyright principles to address issues raised by GenAI without first carefully considering the unintended consequences of such an approach.  Kugler \cite{Kugler2024WhoOA} provides a brief overview of the copyright challenges posed by GenAI and how various governments around the world have begun to address them.

Other research has considered the copyright implications of the emerging base model and the subsequent fine-tuning paradigm \cite{yew2024break} and has studied the applicability of fair use for foundation models \cite{henderson2023foundation}. Additional research \cite{neel2023privacy} presents a review of the discourse surrounding copyright and privacy issues related to LLMs. The research most closely related to our study was conducted by Liang \etal~\cite{liang2024large}, who investigated why developers choose to use AI programming assistants and the challenges they face.
\looseness=-1

Our work builds on this related work by contributing an in-depth exploration of developers' perceptions of the copyright issues surrounding the use of AI-based code-generation tools, both in the U.S. and elsewhere around the world. Understanding these perspectives allows future regulation in this area to be informed by beliefs on the ground rather than in a vacuum. The U.S. Copyright Office, to ``inform the Office's study and help assess whether legislative or regulatory steps in this area are warranted,'' has actively sought public comment on copyright issues pertaining to artificial intelligence~\cite{copyright_office_NOI}. Even after receiving over 10,000 comments, the Copyright Office continues to study this developing area~\cite{copyright_office_1, copyright_office_2}. In the same spirit, our work aims to capture the practices and views of developers to provide a resource for policymakers.

\section{Study Methodology}
\label{sec:design}

The \emph{purpose} of this study is to investigate developers' perceptions about licensing and copyright issues related to the use of GenAI technology for software development. (Although we use the term ``software development'' throughout the article, our study focuses on the use of GenAI for code generation as the main task that can lead to copyright and licensing issues.) 
The primary \emph{quality focus} is to provide lawmakers, regulators, and other interested parties with a summary and analysis of relevant developer perspectives related to the copyright and licensing questions that arise when developing software with the help of GenAI. The \emph{context} consists of 574 developers of open-source projects hosted on GitHub, whom we surveyed via an online survey and follow-up interviews.
\looseness=-1

The study addresses the following research questions (RQs):

\begin{enumerate}[label=\textbf{\labelitemi \space RQ$_\arabic*$:}, ref=\textbf{RQ$_\arabic*$}, wide, labelindent=5pt, leftmargin=5pt]\setlength{\itemsep}{0.2em}
    \item \label{rq:1}{\textit{How do developers use GenAI technologies to develop software?}}  %
    Studying \emph{how} generative AI is used by developers allows us to put the answers provided by developers in response to the subsequent RQs in context.
    \looseness=-1
    
    \item \label{rq:2}{%
    \textit{What are developers' perceptions regarding the licensing and copyright issues that arise from the use of GenAI tools?}} This RQ analyzes several dimensions of the investigated phenomenon, including developers' perceptions of potential emerging legal issues, copyrightable subject matter, and copyright ownership of generated code.
    \item \label{rq:3}{\textit{What other legal concerns do developers anticipate as the use of GenAI increases?}} There are many other potential legal issues to consider as GenAI becomes more widely adopted, including data privacy, malicious content generation, and tort liability.  In this RQ, we report developers' perspectives on these topics, which extend beyond copyright law.
\end{enumerate}

\begin{figure}[t]
\centering
\includegraphics[width=\linewidth]{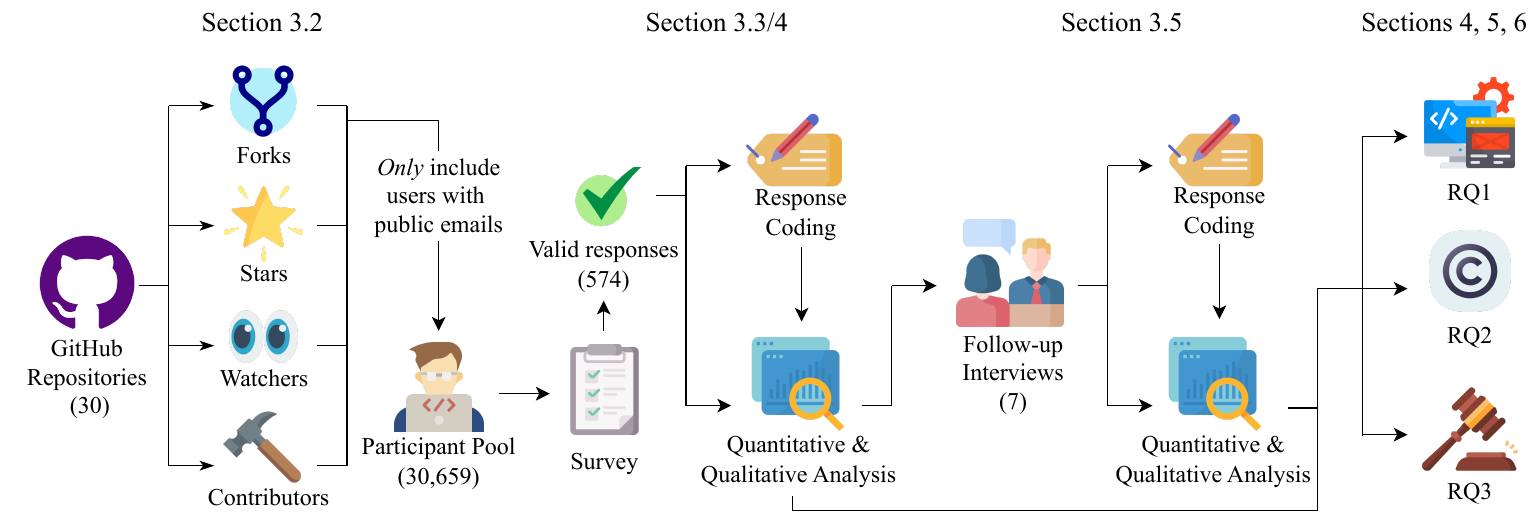}
\caption{Overview of the study methodology}%
\label{fig:methods}
\end{figure}

 Our study, including the survey questionnaire, the participant identification procedure, and survey and interview protocols, was approved by our institution's ethical review board. An overview of our methodology can be seen in \Cref{fig:methods}.%

\subsection{Survey design}

\begin{figure}[t]
\centering
\includegraphics[width=\linewidth]{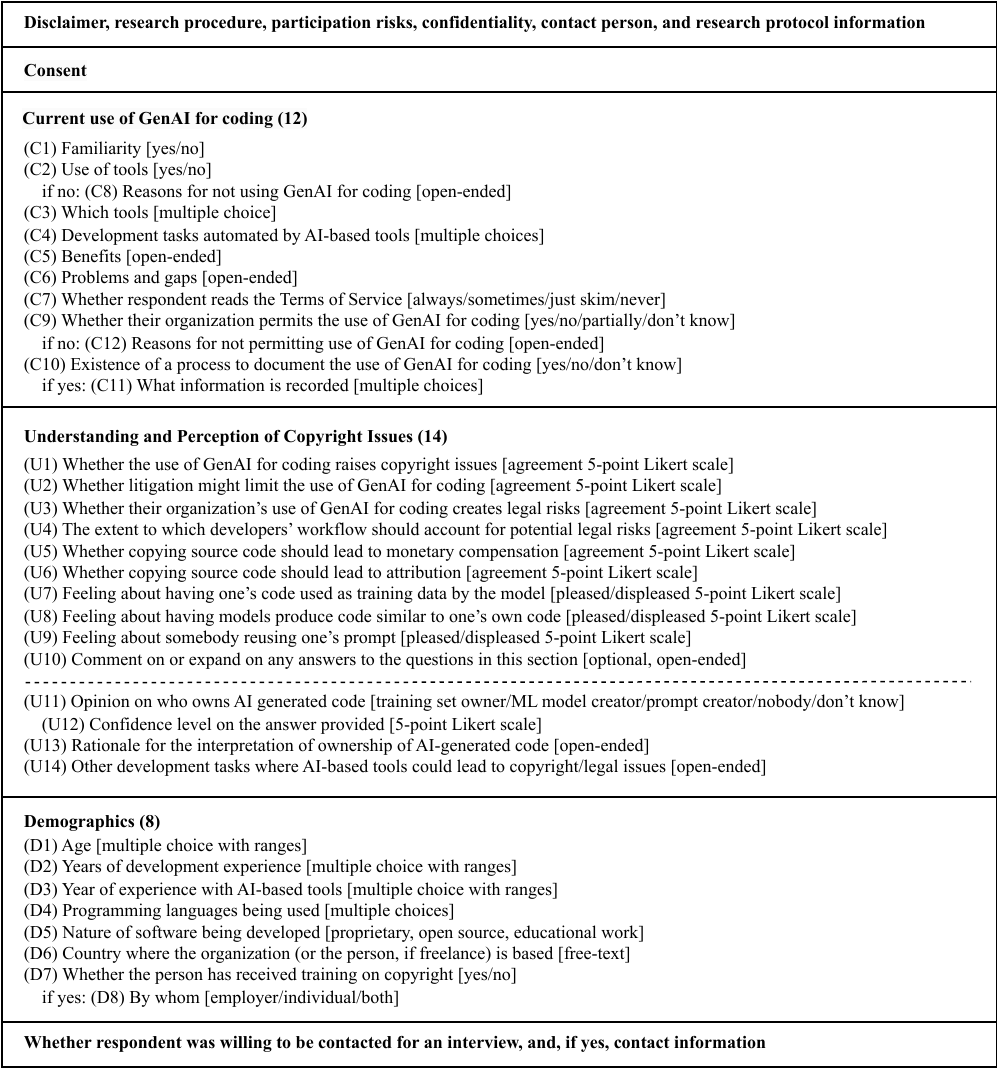}
\caption{Survey design overview}
\label{fig:survey_overview}
\end{figure}

To design our survey, we followed general guidelines~\cite{survey} and SE-specific best practices \cite{DBLP:journals/sigsoft/PfleegerK01,DBLP:journals/sigsoft/KitchenhamP02,DBLP:journals/sigsoft/KitchenhamP02a,DBLP:journals/sigsoft/KitchenhamP02b,DBLP:journals/sigsoft/KitchenhamP02c,DBLP:journals/sigsoft/KitchenhamP03}.  The final questionnaire went through multiple iterations of review and improvements from six SE researchers and one law researcher, all authors of this article. All researchers contributed to the design of the study, but, in particular, the SE researchers contributed domain experience to the formulation of questions about procedures and practices relevant to developers, and the law researcher facilitated the identification of key areas that could be of concern related to copyright and licensing. We carefully formulated the survey questions and ensured that they were written clearly and concisely to avoid biasing the respondents. We also conducted a pilot study with graduate students from our research lab who actively use GenAI tools. Based on their feedback, we further refined our questionnaire, improving questions that were ambiguous or difficult to understand.
\looseness=-1

The survey was structured into three core sections, as depicted in \Cref{fig:survey_overview}:

\begin{itemize}
    \item The section on ``Current use of GenAI for coding'' (12 questions%
    ) asked about respondents' experience using GenAI tools for code generation (C1, C2, C8), including the tools they used (C3); the development tasks for which they used the tools (C4); the benefits and challenges of using such tools (C5, C6); whether they read the Terms of Service for tools (C7); organizational policies for using such tools (C9, C12); and procedures to document tool usage (C10, C11).
    \item The section on ``Understanding and Perception of Copyright Issues'' (14 questions) asked about developers' perceptions of copyright issues surrounding GenAI, including their views on if and how copyright, litigation, and legal risks factor into the use of GenAI (U1, U2, U3); the effects of these concerns on developers' workflows (U4); whether the use of source code in training models should require monetary compensation and/or attribution (U5, U6); the use of code in model training without permission (U7); having models produce code similar to code they created (U8); reusing without permission a prompt that they created (U9); ownership and copyrightability of generated code (U11, U12, U13); and other, broader concerns about copyright/legal issues stemming from development tasks with AI-based tools (U14). %
    \item  The ``Demographics'' section (8 questions) asked about respondents' age (D1); experience with SE and AI (D2, D3); their primary programming languages (D4); whether they develop open source or proprietary software (D5); their location (D6); and if they have received training on copyright law (D6, D7).
\end{itemize}

The survey was designed to be completed in about 15 to 20 minutes and consisted mainly of multiple-choice questions (26), complemented by open-ended questions (7) that allowed participants to elaborate on their responses.
Because we included logic in our survey to ask only those questions relevant to a participant's indicated experience, not every participant saw every question. For example, if a participant indicated that they were not familiar with the practice of generating source code using GenAI (C1), we did not ask them questions about generating source code (C2-C12). The complete survey text and a brief description are found in our replication package~\cite{anonymous_repo}.

\subsection{Participant identification}

We sought to learn from developers who used AI tools for code generation. To this end, and consistent with previous work~\cite{joblin2017classifying,xia2023empirical,stalnaker2024boms,liang2024large,li2023commit,huang2021leaving,filippova2016effects,jiang2017understanding,moraes2021one,constantino2023perceptions,he2023automating,saroar2023developers,liang2022understanding}, we used GitHub as a source to identify potential participants, following multiple steps. %

We began by searching GitHub to identify repositories related to code generation via AI.  We did this by developing a list of relevant tags, including ``coding-assistant'' and ``ai-code-generator.''  The full list of tags can be found in our replication package~\cite{anonymous_repo}.  This resulted in 27 repositories, including open-source coding assistants such as ``Tabby''~\cite{tabby} and ``cptX'' \cite{cptX} as well as ``full development" solutions such as ``gpt-engineer''~\cite{gptengineer} and ``gpt-pilot''~\cite{gptpilot}. 
We included an additional three repositories we believed would be relevant: Meta's CodeLlama,~\cite{rozière2024code},  GitHub Copilot~\cite{copilot}, and Codeium~\cite{codeium}. This brought our total repository count to 30.
Next, using the GitHub API~\cite{githubAPI}, we sought to identify developers who had shown interest in any of these repositories by forking, starring, watching, or contributing to them. We believe that these actions are highly correlated with familiarity with using GenAI tools for software development tasks, although we also acknowledge that our respondent sample may not be representative of all developers and may overrepresent those who frequently use GenAI tools (see \Cref{sec:threats}). Lastly, to respect the privacy of developers, we filtered the collected information to eliminate users who did not provide \textit{publicly available} contact information on their GitHub profile.  This resulted in 30,659 public-facing email addresses, which we used to contact individuals seeking participation in our research study.  %
All information reported in this paper, including demographic information, was contributed voluntarily by respondents who agreed to participate in response to our request. %

\looseness=-1

\subsection{Survey response collection and analysis}

Survey responses were collected using Qualtrics~\cite{qualtrics}. The survey was kept open for six weeks starting on January 21, 2024. Survey invitations were sent out via email during that period.
\looseness=-1

\begin{table}[]
\caption{Response totals}
\label{tab:responses}
\begin{tabular}{|lr|}
\hline
\textbf{Response Type} & \multicolumn{1}{c|}{\textbf{Count}} \\ \hline
All & 772 \\
\hspace{1ex}Complete & 580 \\
\hspace{2ex}Valid & 574 \\
\hspace{3ex}Familiar with AI tools & 554 \\
\hspace{4ex}Incorporated AI into workflow & 497 \\ \hline
\end{tabular}%
\end{table}

We obtained a total of 772 survey responses, 580 of which were complete, as shown in \Cref{tab:responses}.
We removed five responses written in languages other than English, which would have required reliance on translation tools that might not be reliable for domain-specific tasks \cite{henderson2024rethinking}.
One additional response was removed for providing irrational answers to the open-ended questions.  This left us with 574 complete and valid responses. %

Survey responses to five open-ended questions 
were analyzed through a qualitative coding approach \cite{spencer2009card}.  Two authors, both SE researchers, performed \textit{open-coding} by independently assigning one or more \textit{codes} to each response using a shared spreadsheet and codebook.  Each annotator independently coded all 574 responses, adding new codes to the codebook as necessary.  Once the initial coding was completed, the annotators met to settle disagreements and consolidate the set of codes. The disagreements encountered consisted of differing interpretations of the response, applying the same codes in different contexts, or using broader versus more specific codes. Through this process, all such disagreements were resolved, resulting in a final set of codes, definitions, and assignments for each response. %
Our replication package~\cite{anonymous_repo} contains the final codes and definitions.
\looseness=-1

We did not base our analysis on inter-rater agreements because multiple codes could be assigned to each response, and no list of codes existed before the start of coding (\ie our approach was fundamentally inductive). However, we carefully followed best open-coding practices \cite{spencer2009card}, and we leveraged coders' discussions to ensure the reliability of the results. All of the responses were coded in a single iteration by the two annotators. To mitigate agreement by chance and the introduction of bias, the labels assigned to each answer were reviewed by the annotators, including those without disagreement. The same annotators performed this review so as to be able to provide their understanding and rationale behind the labeling of each response. We avoided defining ``umbrella'' codes by creating complete definitions for each code and sharing them between the annotators during the review, as well as by assessing the appropriateness of codes during reconciliation of the annotators' results such that codes that were too similar could be merged together and codes that were overbroad could be broken into multiple, more specific codes. This was done when the annotators observed multiple codes frequently being used to describe the same statement in a given answer or when a code was being applied to answers with significantly different meanings, respectively.
\looseness=-1
 
Two open-ended questions solicited additional thoughts, comments, or explanations on respondents' answers to a number of different questions. The disparity in responses prevented meaningful analysis with our previous coding method, so we applied a different strategy.  One SE researcher read through all 574 responses for both questions and grouped them into categories and sub-categories based on their content by considering the respondent's other answers as context.  An additional researcher reviewed and validated these categorization decisions, discussing and resolving disagreements when needed.  Lastly, as a separate task, a third author reviewed the quotes extracted for the paper, making sure none were miscategorized, misattributed, or taken out of context. %
For the closed-ended questions, we aggregated results using descriptive statistics and discussed them to resolve any disagreements.  One author manually coded each response for questions that had an ``Other'' option (such as C3 and D4), so that this additional information could be effectively analyzed.  Sections \ref{sec:rq1}, \ref{sec:rq2}, and \ref{sec:rq3} present the results of our analysis.

The law researcher contributed an expert understanding of the legal context relating to software licensing issues. This perspective facilitated the interpretation of the survey results and the discussion of the applicable landscape and legal risks presented by the technology.

Where portions of open-ended questions are quoted, we corrected grammar and spelling errors in participant responses for readability.  Deletions and additions to text were occasionally made for clarity or space-related reasons and are indicated by ellipses or brackets. Responses have been attributed to survey participants using anonymous IDs for traceability (\eg \resp{90}).
\looseness=-1

\subsection{Participant demographics and background}

\begin{table}[]
\begin{center}
\caption{Programming language use reported by developers}%
\label{tab:languages}
\begin{tabular}{|lrlr|}
\hline
\multicolumn{4}{|c|}{\textbf{Top 20 Programming Languages}} \\ \hline
Python* & \multicolumn{1}{r|}{407} & Ruby & 11 \\
JavaScript* & \multicolumn{1}{r|}{327} & Bash & 10 \\
C / C++* & \multicolumn{1}{r|}{161} & Kotlin & 9 \\
Java* & \multicolumn{1}{r|}{118} & Swift & 9 \\
Golang* & \multicolumn{1}{r|}{106} & Dart & 6 \\
PHP* & \multicolumn{1}{r|}{84} & Julia & 5 \\
C\#* & \multicolumn{1}{r|}{77} & Shell & 4 \\
Rust & \multicolumn{1}{r|}{55} & Scala & 3 \\
R* & \multicolumn{1}{r|}{41} & Elixir & 3 \\
TypeScript & \multicolumn{1}{r|}{25} & Lua & 3 \\ \hline
\end{tabular}%
{\\ \footnotesize *Language was listed as a multi-select option}
\end{center}
\end{table}

\begin{figure}[t]
    \centering
    \begin{minipage}{0.5\textwidth}
        \centering
        
\begin{tabular}{|lrr|}
\hline
\textbf{Country} & \textbf{Count} & \textbf{Perc.} \\ \hline
United States & 176 & 30.66\% \\
China & 49 & 8.54\% \\
India & 28 & 4.80\% \\
Germany & 26 & 4.53\% \\
United Kingdom & 26 & 4.53\% \\
France & 24 & 4.18\% \\
Brazil & 20 & 3.48\% \\
Canada & 20 & 3.48\% \\
Italy & 14 & 2.44\% \\
South Korea & 13 & 2.26\% \\\hline
\end{tabular}%
\captionof{table}{Top 10 country breakdown*}
\label{tab:countries}

    \end{minipage}\hfill
    \begin{minipage}{0.5\textwidth}
        \centering
        
\begin{tabular}{|lrr|}
\hline
\textbf{Continent} & \textbf{Count} & \multicolumn{1}{c|}{\textbf{Perc.}} \\ \hline
North America      & 198            & 34.49\%                             \\
Europe             & 163            & 28.4\%                              \\
Asia               & 159            & 27.7\%                              \\
Africa             & 27             & 4.7\%                               \\
South America      & 21             & 3.66\%                              \\
Oceania            & 15             & 2.61\%                              \\
Unknown            & 13             & 2.26\%                              \\ \hline
\end{tabular}
\captionof{table}{Continent of Origin*}
\label{fig:dev_origins}  

    \end{minipage}
    {\footnotesize *Some respondents reported working in more than one country/continent}
\end{figure}

Respondents self-selected the nature of their development work from a list that included proprietary software development (331), open source (247), and academia (117). The total number of responses to this question is greater than the number of complete and valid responses (574) because respondents could select more than one option. Respondents came from six continents, as indicated in \Cref{fig:dev_origins}, and 73 different countries (as self-reported), the top ten of which are shown in \Cref{tab:countries}.

Of the 574 complete and valid responses we obtained, 554 developers (96.5\%) indicated that they were generally familiar with using GenAI for producing source code. Of those, 497 (89.7\%) had incorporated GenAI tools into their development workflow, as indicated in \Cref{tab:responses}.  Respondents also used a wide array of programming languages, 48 in total. The most popular languages used were Python, JavaScript, C/C++, and Java, as indicated in \Cref{tab:languages}. Respondents had experience with many different AI tools, including GitHub Copilot (73.6\%), ChatGPT (84.9\%), Bard/Gemini (19.1\%), and Claude (10.3\%), as well as several open-source models (7.8\%); more than 60 tools and models were reported in total, the top 20 of which are indicated in \Cref{tab:ai_tools}.

\begin{table}[]
\caption{GenAI tools reported by developers}
\label{tab:ai_tools}
\begin{tabular}{|lrlr|}
\hline
\multicolumn{4}{|c|}{\textbf{Top 20 Reported AI Tools}} \\ \hline
GitHub Copilot* & \multicolumn{1}{r|}{366} & Phind & 10 \\
ChatGPT-4* & \multicolumn{1}{r|}{304} & Tabnine & 10 \\
ChatGPT-3.5* & \multicolumn{1}{r|}{268} & Mistral & 9 \\
Bard* & \multicolumn{1}{r|}{95} & Ollama & 7 \\
Claude (Anthropic) (any version)* & \multicolumn{1}{r|}{51} & JetBrains AI & 7 \\
Amazon CodeWhisperer* & \multicolumn{1}{r|}{33} & Llama / Llama 2 & 7 \\
Open-source models generally & \multicolumn{1}{r|}{19} & DeepSeek Coder & 6 \\
Codeium & \multicolumn{1}{r|}{16} & Mixtral & 6 \\
Perplexity AI & \multicolumn{1}{r|}{12} & In-house or custom model & 6 \\
Codellama & \multicolumn{1}{r|}{11} & Codium & 5 \\ \hline
\end{tabular}%
\begin{center}
\footnotesize *Tool/model was included in a list of choices; all others were provided \\by respondents to elaborate on their choice of ``Other''
\end{center}
\end{table}

\subsection{Follow-up Interviews}
\label{sec:interview}
To supplement the survey results, we conducted follow-up interviews.  These interviews were designed to establish additional context, obtain answers to questions that arose during the analysis and aggregation of survey responses, and perform a more in-depth exploration of developers' perceptions.  
\looseness=-1

All survey participants were asked at the conclusion of the survey if they would be willing to be contacted for a follow-up interview. Those who consented by supplying their email address (382, 66.5\%) made up our potential interview participant pool. Two authors then selected 22 potential interview candidates from this pool based on various criteria, including experience, likely legal jurisdiction based on location, and indicated views/perceptions on key legal issues, with the goal of interviewing participants representing a diversity of views. The full list of interview selection criteria can be found in our replication package \cite{anonymous_repo}.
 Of the 22 candidates, seven responded to schedule an interview.  The seven interviewees represented four continents (North America (3), Europe (2), South America (1), and Africa (1)) and a variety of backgrounds in software development, including working as a developer of proprietary software (6), contributing to OSS (4), and working in academia (1).

The duration of the interviews was between 20 and 30 minutes. The interviews were conducted using the Zoom video conferencing platform.  Interview sessions were recorded and transcribed using OpenAI's Whisper large-v3 model~\cite{radford2023robust,whisperRepo} to facilitate conversation analysis.  A set of shared interview questions was iteratively derived through the collaboration of six SE researchers and one law researcher. This list of questions is available in our replication package \cite{anonymous_repo}.  Researchers were also free to ask follow-up or clarification questions during the interviews. %

Each interview was conducted jointly by two authors, both SE researchers. The same researchers conducted all of the interviews. The recording and transcript for each interview were independently reviewed by one of the two interviewers to ensure the accuracy of the transcript and extract the relevant portions of each response, assigning topic labels to each one. These quotes and topic labels were hosted in a shared document accessible to both researchers, allowing the reuse of topic labels. Each researcher analyzed roughly half of the interviews. Afterward, the two researchers jointly reviewed the labels assigned to each response to verify the accuracy and completeness of the analysis, discussing and reaching a consensus on the framing and application of topic labels. A third researcher reviewed the quotes extracted for the paper, making sure none were miscategorized, misattributed, or taken out of context.  %
For simplicity and clarity, when we refer to quotes from follow-up interviews, we use the notation \emph{RI} and describe participants by their assigned survey IDs.
\looseness=-1

Interview participants \iresp{31}, \iresp{118}, and \iresp{190} have experience in proprietary software development. (\iresp{118} currently works as a cybersecurity manager.)  \iresp{516} has contributed to open-source projects. \iresp{39}, \iresp{149}, and \iresp{431} have experience in proprietary software development but have also contributed to open-source projects.  \iresp{431} also has some experience writing software in an academic environment.

\section{\ref{rq:1}: Use of Generative AI technologies for software development}
\label{sec:rq1}

\ref{rq:1} aimed to discover how developers are using GenAI technologies to develop software. Although the developers surveyed had a range of perceptions about AI and copyright issues, most seemed to share the belief that the use of GenAI is now a regular part of everyday development activities.
In the following, we present and discuss the findings for this RQ. 
We present the key conclusions that answer this research question as Findings 1-10. A summary of key findings can be found in our replication package \cite{anonymous_repo}. %

\subsection{Uses, benefits, and challenges of GenAI}

In this section, we consider the usage scenarios identified by developers, the benefits developers have derived from this use, and the challenges that developers face.
\looseness=-1

\subsubsection{GenAI usage in software development}

\begin{figure}[t]
\centering
\includegraphics[width=\linewidth]{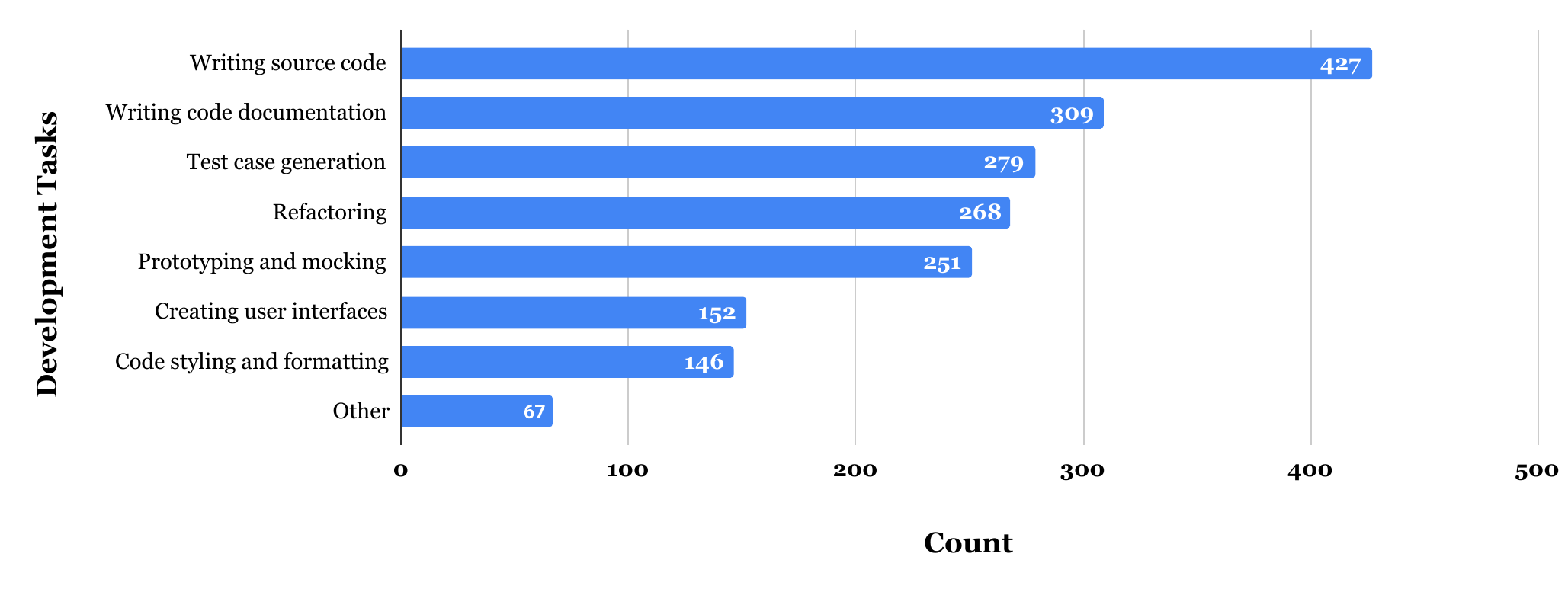}
\caption{Software development tasks where developers reported using GenAI}%
\label{fig:dev_tasks}
\end{figure}

Almost all respondents (554) stated that they were generally familiar with the use of GenAI to produce source code.  Of these, the vast majority---497 (90\%)---indicated that they had incorporated such tools into their development workflows in some capacity. (\Cref{fig:dev_tasks} displays the types of tasks mentioned by respondents.) The remaining 57 (10\%) offered several reasons for not regularly using GenAI, including legal risk (8), personal preference (8), the lack of significant use cases that warrant its use (6), reluctance to send sensitive information to a remote server (6), constraints imposed by an employer (6), and limited trust in the tooling (5). 
\looseness=-1

Among those who had a personal preference against using GenAI, some developers conveyed through open-ended responses the sense that GenAI encourages monotony or lack of effort, despite the fact that GenAI is often used to automate repetitive and boring tasks, as previous work has indicated \cite{abs-2406-07765}.
\resp{188} stated, “I like to think for myself. [W]ithout [that], coding becomes boring and a lot of reading instead of reading and writing.” \resp{339} put it this way: “It's better to write code on my own than to review code generated by AI.” 
The full list of reasons put forward by developers can be found in our replication package \cite{anonymous_repo}.
\looseness=-1

\finding{While the vast majority of developers we surveyed (90\%) reported using GenAI tools, there were many reasons why some developers reported not incorporating GenAI into their development workflows, including perceived legal issues, personal preference, information security concerns, and limitations placed by employers.}

Of those familiar with GenAI tools, relatively few developers (32) %
said that their organization did not permit the use of GenAI for coding. These developers reported their understanding of the restriction.  %
Twelve (38\%) stated that such usage could compromise sensitive or proprietary information. Ten (31\%) indicated that their organizations were concerned about the legal implications of using such tools. Four (13\%) suggested that their employers were slow to adopt new technology and did not fully understand its benefits.  Three respondents (9\%) pointed to the fear of security issues, such as incorporating insecure code or designs. Notably, these responses could reflect overlapping concerns: compromising proprietary information, for example, could also raise legal concerns.

\finding{Of the relatively few respondents who reported working at organizations that disallowed the use of GenAI, about a third indicated that they believed the restriction was based on fear of proprietary information leakage and the potential legal challenges that such usage might bring.  Some developers also perceived that their employers were slow to adopt new technology.}

\subsubsection{GenAI tools used by developers}
Participants used various GenAI tools, including open-source, open-weight, and proprietary models.  In total, developers reported using 64 distinct GenAI tools, of which the top 20 can be seen in \Cref{tab:ai_tools}.  Developers were encouraged to identify all the tools they had used in their work. Of these, the top six (which were also provided in a list of choices, along with an ``Other'' option) were: GitHub Copilot (366), ChatGPT-4 (304), ChatGPT-3.5 (268), Bard (95), Claude (51) and Amazon CodeWhisperer (33). The next largest group (19) reported that they used open-source/open-weight models, with several such tools/models listed by multiple respondents, including Codellama (11), Mistral (9), Ollama (7), Llama2 (7) and Mixtral (6). %
Developers also used other proprietary models and services, including Codeium (16), Perplexity (12), Phind (10), and JetBrains AI (7), and six respondents indicated that they used custom-trained or internal models for their development work.  A full list of the tools that developers reported using can be found in our replication package \cite{anonymous_repo}.  

\begin{figure}[t]
\centering
\includegraphics[width=.8\linewidth]{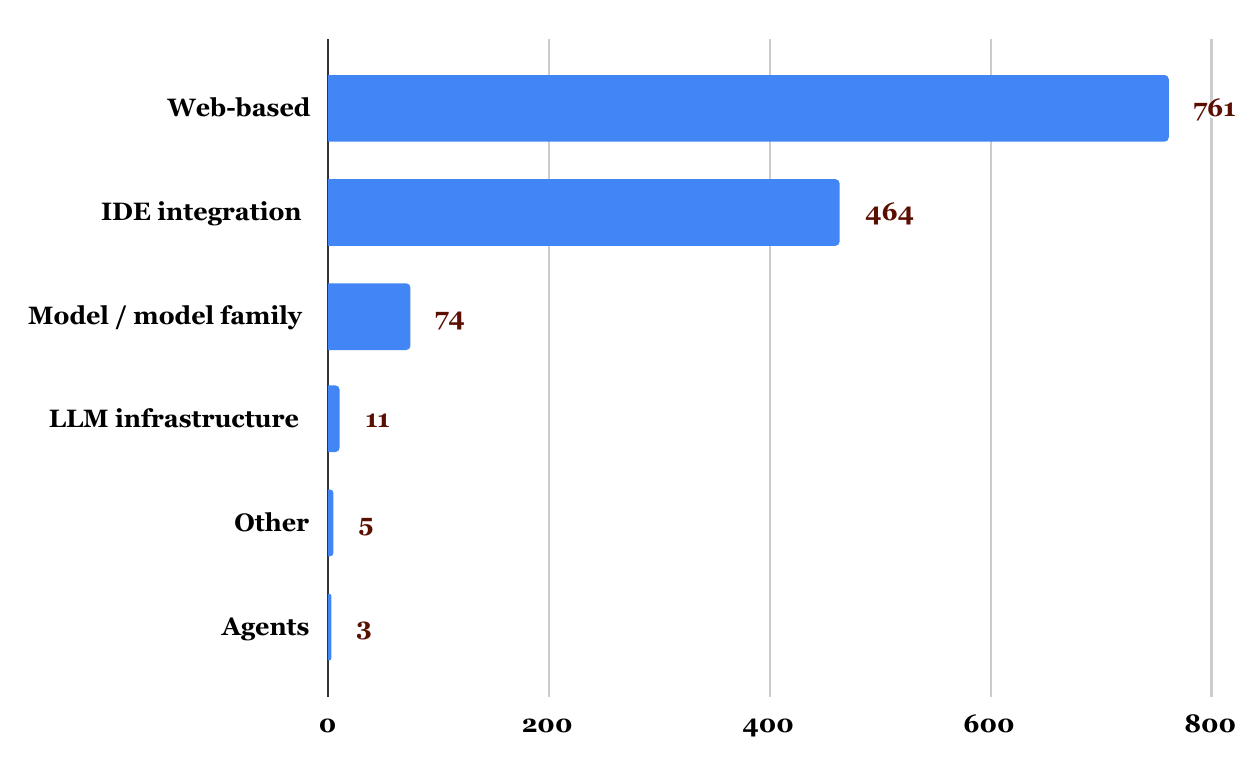}
\caption{Instances of tool category uses reported by developers}%
\label{fig:tool_groups}
\end{figure}

Of the tools listed by respondents, 20 were designed to integrate into an IDE.  Many of these tools provide autocomplete suggestions for developers based on the context of surrounding code, but some also include Q/A chat functionality. Developers also reported using 18 distinct web-based tools, such as ChatGPT, Perplexity, %
and Hugging Face Chat~\cite{chatgpt, perplexity, huggingChat}, which require the user to provide more context than IDE-integrated tools.  Respondents also indicated using four model infrastructures: Ollama,  Oobabooga, llama.cpp, and Mozilla llamafile~\cite{ollama, oobabooga, llamacpp, llamafile}.  These infrastructures provide a streamlined means to run and train open-weight models locally. In total, respondents identified 15 distinct open-weight models/model families. (We note that the prevalence of open-weight models and open-source tools in this space has probably increased since our initial survey with the subsequent release of models such as Llama3~\cite{llama3_release} and the open-source of XAI Grok~\cite{xai_grok}.) Three developers also noted the use of GenAI agents like gpt-engineer~\cite{gptengineer} and metagpt~\cite{hong2023metagpt}, which allow models to self-prompt to accomplish more complex tasks. Lastly, four GenAI tools that were not designed for software development tasks, %
such as Figma~\cite{figma} and GrammarlyGo~\cite{grammarly}, were mentioned. \Cref{fig:tool_groups} shows the usage of tools in the categories discussed here as reported by respondents. 

 \finding{The ecosystem of GenAI tools developers reported using was varied and extensive, comprising both open-source/open-weight and proprietary solutions.  Developers primarily used web-based tools and those that could be integrated into their IDEs.}

\subsubsection{Benefits derived from GenAI usage}
Respondents who had used GenAI as part of their workflow identified several benefits from its use.  The greatest benefit mentioned by 288 developers~(58\%) was increased productivity, faster development, and efficiency. Other benefits included not having to spend time writing simple or boilerplate code (88), quick prototyping (42), debugging (37), faster, more accurate documentation (29), summarizing existing documentation as an alternative to reading it (28), providing inspiration (16), refactoring (14), use in writing test cases (5), domain logic support (4), and generating test data (4).  A full list of the benefits identified by respondents can be found in our replication package \cite{anonymous_repo}. We note that participants often mentioned tasks as benefits, likely indicating that GenAI improved the effectiveness, productivity, or ease of completion for those particular tasks.

Thirty respondents identified improvements in code quality and optimization over what human developers could achieve. \iresp{118} noted, ``When you couple the lack of training [of the average developer] with generally one to three years' experience, [...] the fact is that GPT4 or many of these tools are actually better developers than they are. [...] I would almost feel better if [a code comment] said I wrote this and then got a code review from GPT4.'' 
Similarly, developers reported that GenAI could help them write software in unfamiliar languages~(28) and libraries (27). During a follow-up interview, \iresp{190} described how they would ``basically just writ[e] the code in Typescript and then just tell [ChatGPT] convert to Liquid.''

Six respondents mentioned using GenAI specifically to port or translate code.
\resp{90} summarized this benefit: ``The syntax of various languages is [why it] can take hundreds of hours of use to establish [familiarity] in a new language, but AI significantly reduces the ramp-up time.'' Along the same lines, nineteen respondents noted that GenAI helped them become better developers by improving their understanding of unfamiliar code.  %
However, GenAI models are not limited to explaining source code or concepts. \resp{238} noted that they could also be used for ``commit message generation [and ex]planation of changes in commit (summarization).''

 \finding{The benefits developers reported from using GenAI tools for coding go beyond increasing productivity and avoiding repetitive, boring tasks. They also included a better understanding of code, improved code quality, and code optimization, which some human developers might struggle to achieve.}

\subsubsection{Challenges and shortcomings encountered in using GenAI}
Respondents also identified several challenges that arise from the use of GenAI, the most significant being the generation of unhelpful, unwanted, or broken code (128). Additional challenges included hallucinations (64) (such as recommending nonexistent libraries or method calls), problems arising from outdated training data (53), issues arising from tools not having access to the project's wider context (52), which could be caused by insufficient context windows (40)%
, and an inability to use GenAI to solve complex problems (43). %
\resp{148} elaborated: ``AI can do simple code pretty well, it's not that capable [of generating] more complex code, it's incapable of generating well-architected code outside of extremely simple patterns.'' \resp{33} provided an example: ``In my experience, a simple Python/Django project that required some form customizations was just not possible for any AI tool at my disposal. The logical part, the thinking, and the experience gained from years of coding real-world applications and systems design are [not comparable to] an AI code generation tool like GPT or Gemini.''  The full list of challenges reported by developers is included in our replication package \cite{anonymous_repo}.

Twenty-five respondents noted that the ability to write good prompts is essential to effectively using GenAI tools.  \resp{303} reported that ``AI cannot complete the majority of tasks by itself, even simple ones[,] without time spent [by the developer on] prompt engineering.'' 
Writing effective prompts is not always an easy task, however. \resp{18} said, ``It is hard to communicate what I want or what I'm trying to describe to the model.''  
If done correctly, however, prompting was also seen as a way to overcome other limitations of the technology.  \resp{240} said, ``There is a limit to the complexity that can be generated stand[ing] alone, but a human crafting intelligent prompts can still architect complex solutions using AI.''

\finding{Aside from obvious problems such as possible hallucinations, participants reported that the use of GenAI for coding presents challenges related to the lack of suitable contextual information available to the model at inference time, the inability of such models to solve complex, non-trivial tasks, and the need to provide good, well-designed prompts.}

\subsection{Current practices for documentation and legal compliance when using an AI tool}%

\subsubsection{AI-generated code review and compliance}
Given the potential challenges associated with using GenAI, strategies to mitigate security and compliance risks become increasingly important.  (Those strategies can differ depending on the environment. \Cref{sec:interview} provides demographic information on participants, including development experience background.)

As \iresp{118} noted, when code is not generated by the developer, review can be a challenge: ``[T]hat's another big part of it is just going through line by line, but it's like a more passive mindset, which I find a bit challenging compared to actually writing the code, so it can be harder to actually spot what's wrong with it.'' Two interview respondents, accordingly, believed that AI itself could assist with reviewing AI-generated code. \iresp{149} told us that they ``ask the GenAI itself to review the code, perhaps using a different model [...] [and] ask it to give me gotchas or things to look out for.'' \iresp{31} tentatively agreed but acknowledged the challenges: ``I've thought about this [...,] and [if] it's something where the answer is using machine learning to check whether or not the code that you've used appears, and would put you at legal risk, [...] you're relying on a large language model to check whether your large language model puts you at legal risk.''
 
Two interview respondents reported employing manual checks on sites like GitHub or Stack Overflow to determine the provenance of AI-generated code to assess, for example, whether they are in compliance with that code's license. \iresp{431} emphasized the manual nature of their compliance review: ``I don't really have any tool [to verify provenance], so that's why I'm doing this step of going to a website such as Stack Overflow and trying to see if the code which has been given to me by the model has already been proposed somewhere else on the Internet.'' But \iresp{516} noted that these manual processes are limited, because ``if it's not open source, I probably have no way to or means to check it because closed source is not generally available [...] in a search or anything like that.''

Other developers reported an overall review that did not focus on vulnerabilities, and still others stated that they avoided compliance checks on AI-generated code altogether for fear of opening themselves or their organizations to liability. \iresp{39} explained, ``I have gotten advice from IP lawyers at a previous job [...] to not look at patents for anything that I'm working on. Because if there is evidence that I have seen a patent that we then infringe on, then it's willful infringement, which can mean triple damages or something like that. [...] I could imagine a case where it might make sense to not track down which lines of code were generated by a LLM, because then you might have less of a willful infringement case or something like that.''

\finding{Respondents reported that their GenAI code review is typically done manually or by using other GenAI to check for code compliance. Some developers reported concerns that the process of conducting a review may itself expose them or their employer to liability.}

\subsubsection{Documentation of GenAI usage}

\begin{figure}[t]
\centering
\includegraphics[width=\linewidth]{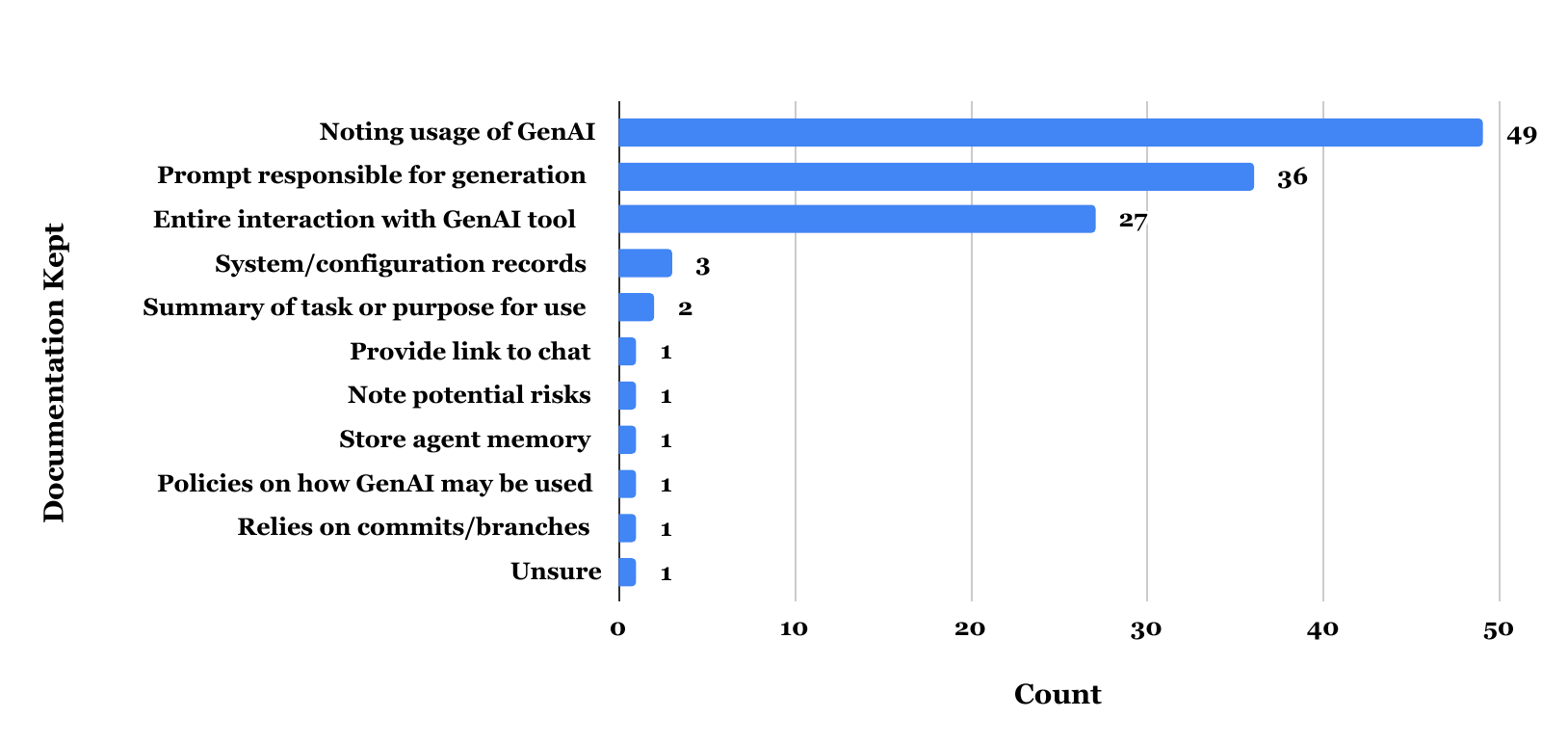}
\caption{Information collected/stored during documentation processes}
\label{fig:documentation_strategies}
\end{figure}

One might expect that organizations that permit the use of GenAI for code development would ensure that developers were aware of any processes to document its use so that the legal risk to the organization could be assessed by appropriate personnel.  However, of the 445 developers who indicated that their organization permitted or partially permitted the use of GenAI, only 83 (19\%) indicated that they were aware of any documentation process, and of the remaining group, 312 (70\%) said that there was no process and another 50 (11\%) were uncertain if a process existed. \resp{325} offered one view of why a process was unnecessary: ``I did not use anything that cannot be [G]oogled anyway---but we never thought about having to document or question code we found through Google.'' Whether this means that many organizations have no process for documenting the use of GenAI or that at least some developers incorrectly believe that no documentation process exists at their organization, the result may be that developers do not have a full understanding of the possible legal risks they or their organizations face (as described in \Cref{sec:legal_background}).

\finding{\label{finding:no_documentation}Most developers in our study (81\%) reported that there was no process for documenting GenAI usage within their organization or that they were unaware of any such process.  To the extent that organizations have processes in place to document the use of GenAI, they may not be adequately educating their employees about these processes.}

We further asked the 83 %
respondents who reported that their companies had processes to document their AI usage on what information was collected during the process. (Respondents could select multiple options and provide additional items; the results are presented in Figure 5.) The majority of the responses (49, 59\%) indicated that documentation consisted of noting the use of an AI tool, and 36 respondents (43\%) indicated that the prompt was also documented. In 27 cases (33\%), the entire interaction with the code generation tool was documented. Other documentation strategies included keeping system/configuration records (3), summarizing the purpose of using AI tooling (2), providing links to chats (1), documenting known risks associated with the use of AI (1), storing the AI agent's memory (1), and relying on version control (1).

\finding{%
\label{finding:documentation_process}The most common documentation process respondents reported was to note the fact of GenAI usage and/or to document the prompt used to acquire the generation. However, almost a third of the respondents whose companies had documentation processes (33\%) indicated that they documented the entire interaction, potentially leading to reproducibility.}

One motivating factor behind documentation was to explain the cause of a potential failure rather than to provide provenance. \resp{190} noted, ``I usually write a comment `written by [C]hatGPT' or `partly written by [C]hatGPT' to denote that if it breaks, I would maybe have written it better but used [C]hatGPT to speed up the process.'' 
\iresp{118} suggested, by contrast, that the typical training at their place of employment does not anticipate this type of activity: ``[Developers are] allowed to start using the tools and there's guidance on where and when they can't, but not in terms of attribution or [adding] warning[s that say,] `Hey, here be code generated monsters.{\rq}{\rq\rq}

\subsubsection{Review of the Terms of Service for AI Tools}
\begin{figure}[t]
\centering
\includegraphics[width=.85\linewidth]{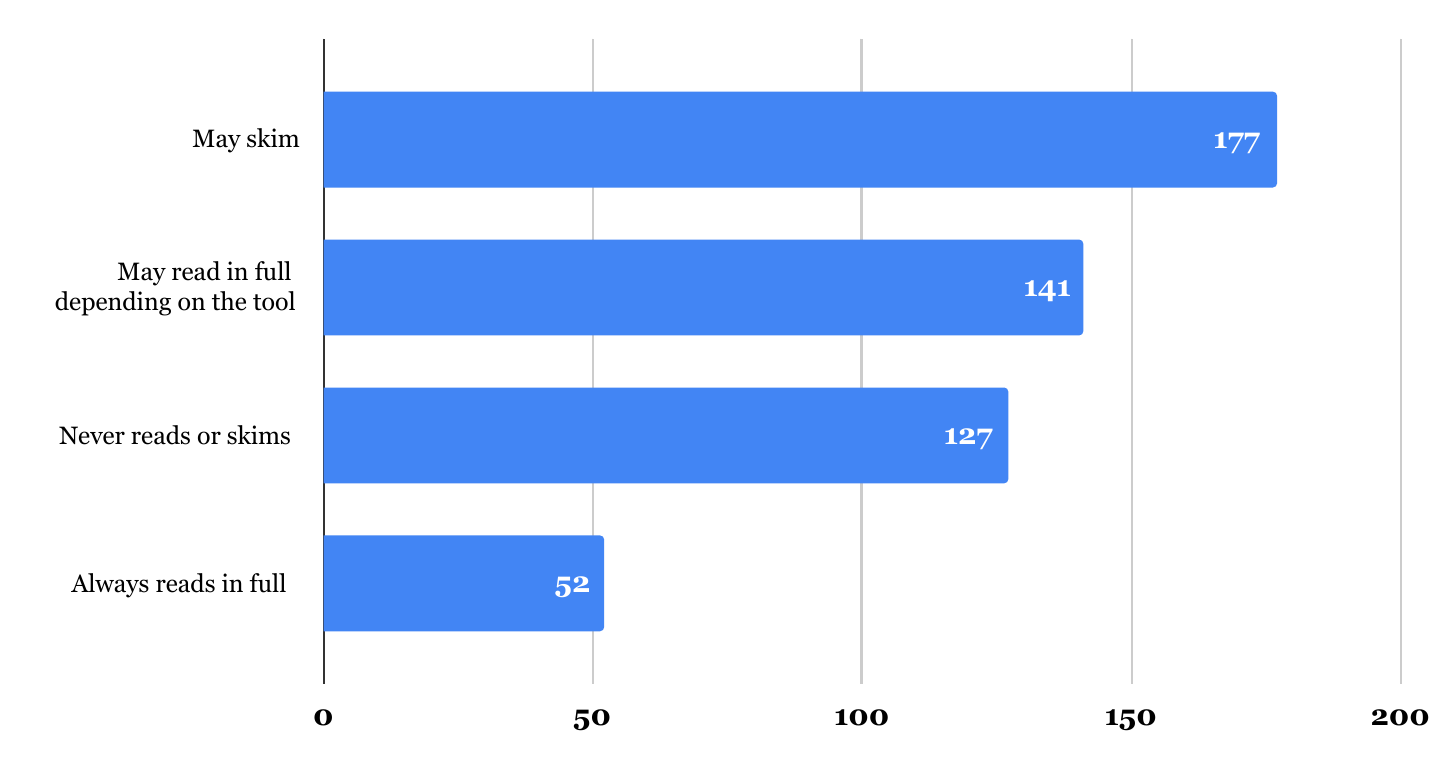}
\caption{Developer approaches to Terms of Service (ToS) for AI Tools}%
\label{fig:terms_of_service}
\end{figure}

The Terms of Service (ToS) for GenAI tools convey how developers can interact with models by describing permissible and impermissible behaviors.  These ToS may contain language that prevents the use of the service for any illegal activity, forbids the use of model output as training data for another competing model, or (for open-weight models) provides guidance on what is allowed when fine-tuning.  The ToS may also provide notice that data provided as user prompts may be used as training data for future models.  In short, it is important for both users and model fine-tuners to understand the terms that they are agreeing to before using a model.

Developers' views on the ToS of GenAI tools varied widely, as reflected in Figure 6. A little over a third of respondents who said they incorporated AI into their workflow (177, 36\%) indicated that they never read terms in full but might skim them looking for relevant or problematic provisions.  A similar percentage of respondents (141, 28\%) stated that they sometimes read the ToS in full, depending on the tool they are using. An additional 127 respondents (26\%) claimed that they never read the ToS for the tools they use.  The smallest group, 52 participants (11\%), told us that they always read the ToS in full.
\looseness=-1

In follow-up interviews, respondents who tended not to read the ToS provided their rationale.  
This decision was often based on factors related to time, with \iresp{431} citing the ``length of the text,'' or having other priorities, as indicated by \iresp{190}, who described having more impactful development work to do than reading the ToS.

Other respondents indicated that they relied on others to read the documents and alert them to relevant provisions. Respondents who worked within organizations stated that they trusted that their organization's legal department would take responsibility for compliance. \iresp{39} stated, ``I personally don't worry about [licensing issues related to generated code] because the legal department where I work has already authorized [the] use of [the LLMs], and it's kind of their problem from my perspective.'' 

Others noted that they relied on discussions on blogs and social media to identify problematic provisions. \iresp{190} stated, ``If it would blow up on the internet that the terms are bad, then I would kind of hear about it.'' 
\iresp{516} referred to the website Terms of Service; Didn't Read (\url{https://tosdr.org}), which ``[does] a summary of the main points of each Terms of Service and privacy policy on the services that have already been analyzed. It's analyzed by a community [...]. The main points of concern [...] are marked in red, so you need to watch out for them.''

The group of respondents who stated that they did read the ToS for the tools they used reported focusing on how the tool developer acquires and uses both personal data and training data, at least for some tools. \iresp{516} explained, ``I always take a look to see what [GenAI tools] do with the data, the risks involved in inadvertently disclosing personal or private information into the data.''
\iresp{31} expressed similar concerns, ``The training set, making sure that the training set is---they haven't trained it on stolen code. And then the other one is making sure that I can run the whole thing locally and maintain privacy.''
Some others, for example, \iresp{190}, took a more pragmatic approach, stating, ``[The major red flags] would have to be something that could bring me or the company I work at into big trouble, and that would [have to] be more trouble than that we probably won't exist in a year because our product isn't good enough, so it would have to be pretty big.''

\finding{Even though some provisions in an AI tool's Terms of Service may be important to developers and their work, developers reported not always thoroughly reading the Terms of Service for the tools they use, choosing instead to skim the document, assume that others in their organization have read the document, or rely on information and validation from others in the development community.}

\subsubsection{Copyright/legal training}
Of the 574 respondents, only 68 (12\%) indicated that they had any formal training in copyright law and/or the legal implications of using code generation models.  Of these, 28 (41\%) sought the training out on their own.  Sixteen respondents (24\%) indicated that the training was provided by their employer, and 24 (35\%) described their training as some combination of self-initiated and employer-provided. (Some respondents interpreted the question to refer to GenAI training more generally rather than training that focused on legal implications.)

In our follow-up interviews, respondents provided further information about the nature of the training reported by developers. \iresp{431} described their training as ``two courses [during my master’s degree about] [...] everything related to the law about anything digital.''
By contrast, \iresp{31} experienced a dearth of resources: ``I started looking for training not just for myself but for our team, and it wasn't something where I could really find a whole lot. [...] There aren't a lot of resources out there for that because everybody's still figuring out the legal implications.''

\finding{%
Relatively few respondents (12\%) reported having undergone any formal training on the legal implications of GenAI. Some struggled to find suitable available resources.}

\section{\ref{rq:2}: Developer perceptions of licensing and copyright issues emerging from the use of Generative AI}
\label{sec:rq2}

\ref{rq:2} sought to determine developers' beliefs and understandings regarding the licensing and copyright issues that arise from using GenAI tools. Here, we explore developers' thoughts on current litigation and the prospect of regulation, sentiments regarding the use of their own code in model training data, opinions on the ownership of GenAI output and prompts, and the perceived risk of copyright infringement. This research question is answered by Findings 11-24, which are summarized in our replication package \cite{anonymous_repo}.

\begin{figure}[t]
\centering
\includegraphics[width=.9\linewidth]{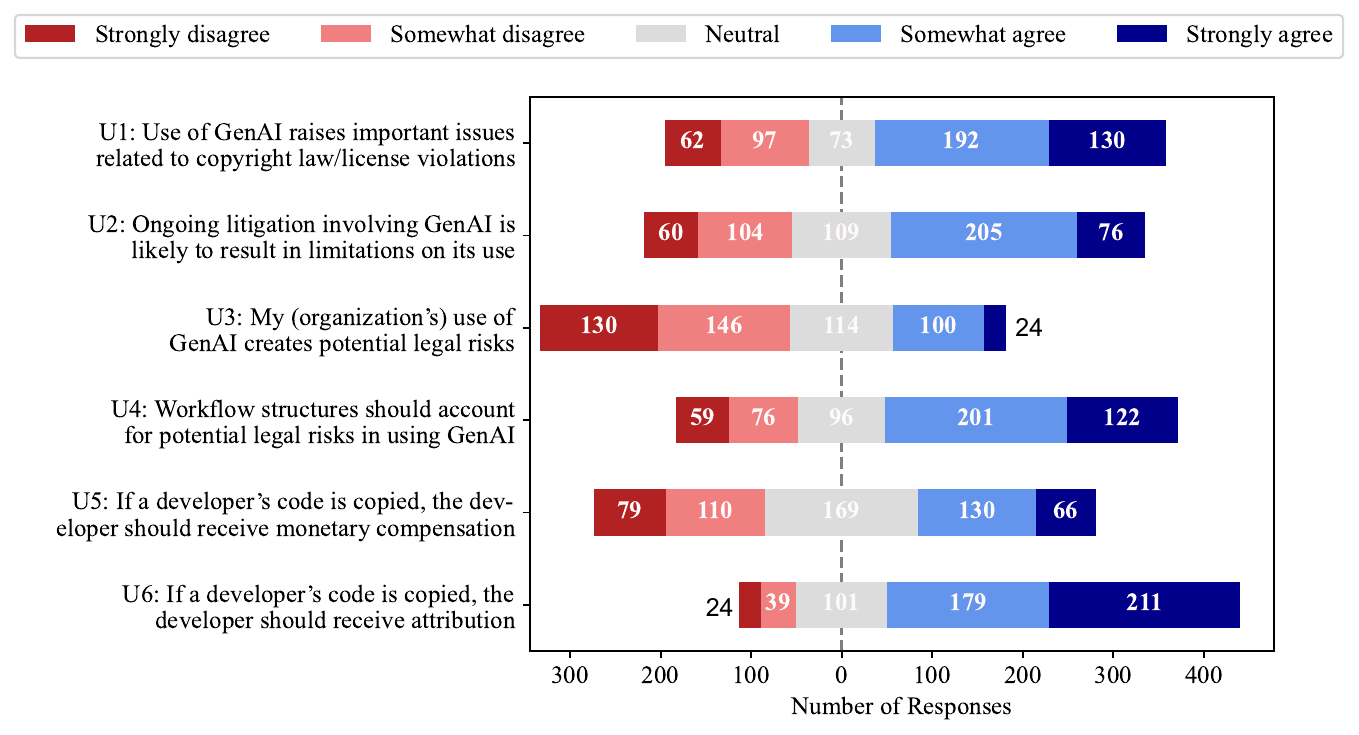}
\caption{Developer perceptions on copyright and GenAI}
\label{fig:likert_1}
\end{figure}

\begin{figure}[t]
\centering
\includegraphics[width=\linewidth]{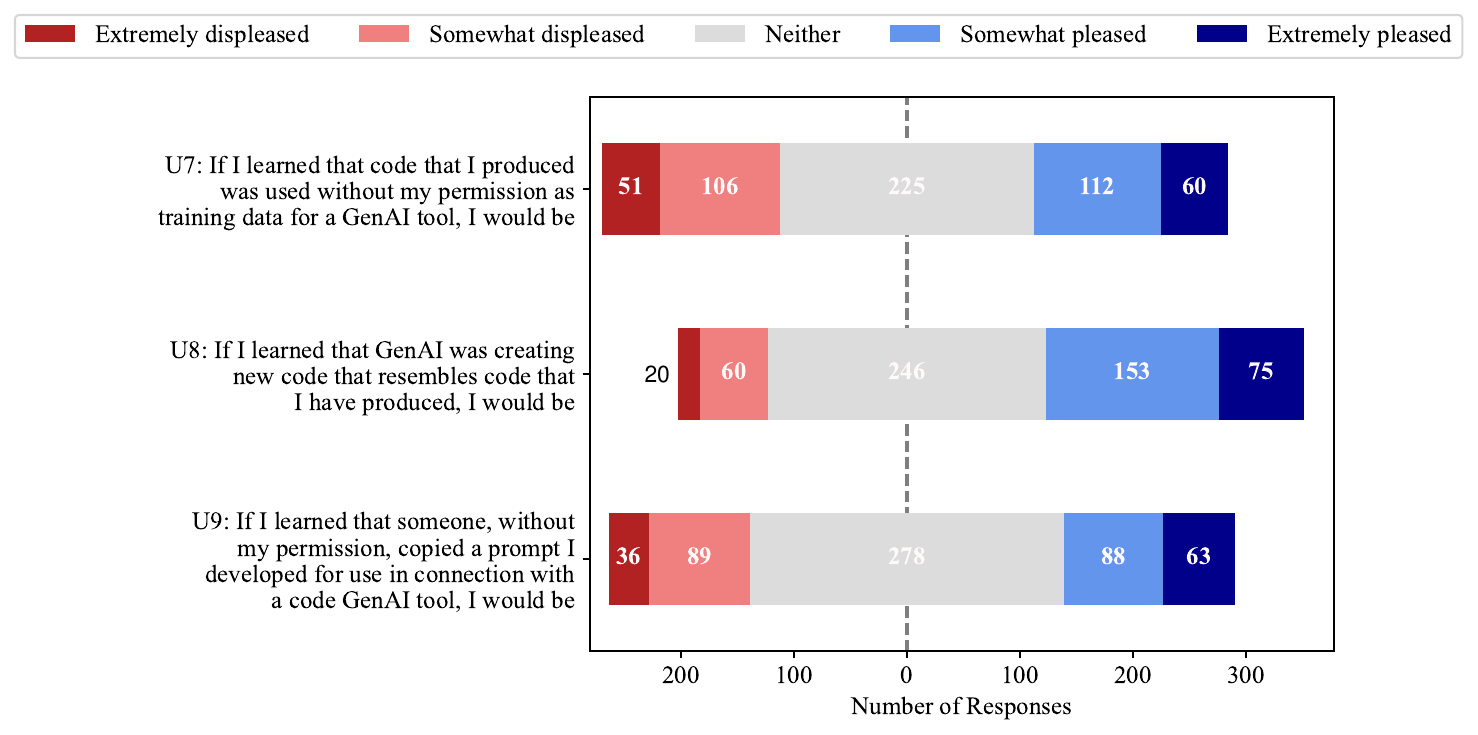}
\caption{Developer attitudes towards copyright and GenAI}
\label{fig:likert_2}
\end{figure}

\subsection{Perceptions about litigation and regulation}

We asked respondents whether they believed that litigation would result in limitations on the use of GenAI for software development. (\Cref{fig:likert_1,fig:likert_2} show respondents' answers to nine five-point Likert-scale~\cite{oppenheim2000questionnaire} questions.) Of the 554 respondents who indicated familiarity with AI tools, 205 (37\%) somewhat agreed with the view that current litigation efforts will result in limitations placed on the use of GenAI models for software development, with 76 (14\%) respondents strongly agreeing.  An additional 109 (20\%) respondents were neutral, and the third largest group of 104 developers (19\%) somewhat disagreed. The smallest group, 60 developers (11\%), strongly disagreed.

Although more respondents believed that litigation would result in limitations on GenAI use, developers who believed that litigation would \textit{not} result in limitations were more vocal about their views in their responses to open-ended follow-up questions. Many such respondents indicated that technology would outpace attempts to regulate it, either through litigation or otherwise. \resp{98} commented, ``I don't see any limitations coming due to ongoing litigation---yes, it may cause issues for OpenAI and StabilityAI, but you can't put the genie back in the lamp.'' \resp{507} further elaborated: ``Pandora's box has been opened. Even if the current models were legally limited, good luck proving that certain code has been used as training data, and good luck stopping people from running Llama instances locally.'' \resp{54} acknowledged the existence of legal issues but believed that ``the momentum that's already been built by OpenAI and all the hype around the tools means there will not be the political capital to get any restrictive regulations in place.''

One respondent (\resp{239}), after weighing the pros and cons of regulation, concluded that no regulation at all was better than regulation that hampered development: ``On the one hand, I know it's sort of a legal gray area right now and ultimately we're going to need legislation. On the other hand, these tools are incredible[,] and I can't imagine going back to not having Copilot. It would be like riding a bicycle with a flat tire. I would like to see sensible legislation that promotes growth and advancement of these technologies, but I'm worried that we won't get that. I'd rather it continue to be a free-for-all---including, yes, commercial models trained on freely available source code---than have the tools taken away or severely restricted.''

\resp{329} recommended a cautious approach to regulation and litigation, stating that if ``[l]awyers get involved before the technology matures[, we are] putting the carts before the horses. Preemptive copyright protection can only hurt technology innovation.''  In the same vein, \resp{185} said, ``I am skeptical about what I consider to be premature efforts to restructure the legal framework around AI until we get a better handle on its implications and potential.''

Finally, %
two respondents expressed that they thought it would not be technology that changed to adapt to the law but rather the law that would change to adapt to technology. \resp{376} commented, ``The result [...] will probably not be limitations on the technology since open-source is rapidly developing and cannot be genuinely regulated in this fashion. Rather, intellectual property, artistic rights, etc. will need to be recalibrated.'' 
\resp{182} offered an aspirational approach, stating that they ``don't think it is wise to approach the question of copyright of AI-generated/assisted code as a matter of whether the generated code is or is not copyrightable based primarily on what has been settled. Rather, we should think about what we want it to be, keeping in mind the benefits [that] copyright is intended to bring to humanity.''

\finding{\label{finding:litigation_and_speed}While about half of the participants anticipated that litigation would result in limitations on GenAI use, a vocal minority suggested that imposing such limitations would be difficult, if not impossible, and others urged restraint, given the pace of technological developments.}

\subsection{Using one's code in models' training data or as output}
If regulation of GenAI occurs, whether through legislation or litigation, the perceptions and practices of stakeholders, including software developers, should be considered.  Our study therefore sought to learn respondents' views on the use of code as training data for generative models.

\subsubsection{Sentiments regarding the inclusion of one's own code in training data}
\label{subsec:sentiments_around_inclusion}

Of the 554 respondents familiar with GenAI tools, the largest proportion (41\%), as seen in \Cref{fig:likert_2}, indicated that they would be neither pleased nor displeased if the code they developed was included in training data without their permission. However, open-ended responses indicated that the reality is context-dependent. If the work was open source, respondents generally indicated that they would be pleased or even flattered to have their code included in training data as a way in which they could contribute to the greater good. \resp{144} expressed this sentiment: ``If it's just a class, or a few elements from my code, I would actually be pleased and proud that my code is having [an] impact on other peoples' work and on the advancement of science in general.'' \resp{67} elaborated, saying, ``If the code has been used [in training data] with my implicit (\eg through BSD-licensed code) or explicit (\eg through my company's or my permission) then I would be happy to have contributed. Otherwise, [my] answer becomes `displeased.'\thinspace'' (Some respondents, however, expressed concerns about training models on open-source code given the quality of the training data. \resp{196} stated, ``[M]ost of the code I write is bad or middling, and the lack of intentional design I've seen in the code of open-source projects [...] really worries me [...] [in that] we can just automatically generate unclear and potentially insecure code.'') 
\looseness=-1

In contrast, developers indicated disappointment about the idea of their proprietary code being included in training data. \resp{82} said, ``I'm perfectly OK if a company trains their model on MIT [licensed] projects that I have on GitHub, but GitHub training their model on [proprietary] code is a bit different.'' Similarly, \resp{98} indicated that for private projects, ``I wouldn't be happy if the code was used for LLM training mostly because some of these repositories may contain committed secrets,'' and for work projects, ``the code in these is in private repositories and everything that's inside is a corporate secret. If that code is used for LLM training, it could harm my employer by exposing secrets and algorithms we use, which would be extremely displeasing.''    %

Consistent with the traditional justification for copyright law in the United States, \resp{148} reasoned that unauthorized use of work as training data for GenAI could disincentivize human creativity, leading to the reluctance of developers to contribute to open source:
``[This] is more of a case of how to sustain OS[S], than [it is] a case of code being copied.'' Indeed, \resp{245} said, if developers are not incentivized to create code, ``we will end up with a gradual decline in the quality of all intellectual property as current AI systems are not creative enough to produce IP that is significantly different from the training material.'' \resp{514} offered a similar view: ``The real issue to me is that those engines are [parasiting off] websites like [S]tack [O]verflow [and] actually lowering the amount of money they make. If these websites stop [...] produc[ing] good answers that [AI tools] can feed [off of], they [the AI tools] all will die.'' \resp{604} even speculated that ``[c]ode generation tools might encourage creative professionals to hoard knowledge, defeating the original purpose of copyright: enabling protected[,] limited-time monetization followed by release to the public domain.''
\looseness=-1

\finding{\label{finding:training_data}Developers’ views on having their code included in models' training data were context-dependent. If the code was open source, many developers were indifferent or even pleased to have their work included; this was not true for proprietary code. Developers also expressed concerns about the quality of the code incorporated in the training data and whether increased use of GenAI tools would disincentivize developers from contributing to the OSS community.}

\subsubsection{Conditions around use of one's code in training data}
\label{subsec:conditions_around_use}
If using another developer's code in training data requires permission, the question then becomes what conditions might be imposed, including monetary compensation and attribution.

Eleven respondents used open-ended questions in the survey to provide their thoughts on whether developers should receive monetary compensation for their code being included in training data. 
Four of these argued that training on open-source software did not necessitate such compensation or, indeed, that the developers of open-source software had waived their right to any compensation. Six identified some conditions under which they considered it appropriate. 
\resp{159}, for example, stated, ``[I]f my code is free, and the AI model is paid, I feel wronged and deserve compensation.'' 
\resp{613} believed that developers should always receive some form of compensation: ``[T]he developers whose code these big corporations use to train their models must receive some amount of compensation in terms of free credits, [GPU] time[,] etc. on their platform.''

\resp{433} took a broader view of the impact of compensation: ``If monetary compensation were required to train these models, it would put smaller AI development companies out of business and restrict the currently thriving sector to only the Tech Giants. No one else would have the [means] to not only acquire the data but also to do the training.'' 

\finding{\label{finding:monetary_compensation}Some respondents suggested that training data contributors should receive monetary compensation or other benefits, such as free access to the trained model or free credits for the platform.  However, requiring AI companies to provide monetary compensation to all contributors to the training data could make it more difficult for smaller AI companies to enter the market.}

Eight respondents stated that it was also important to respect licensing terms, including attribution for their contributions or (for restrictive licenses) reuse under the same terms.  As respondents suggested, the modes of attribution could be as fine-grained as a citation provided along with the generated code or as coarse as a manifest that included links to all the training data. In the view of respondents, embedding attribution into the AI model would serve two functions: (1) respecting the rights of developers and (2) helping users and organizations ensure that they comply with the licenses of code they use by providing provenance of the output. In terms of the latter, \resp{456} imagined a situation in which an organization using GenAI unintentionally reproduced GPL-licensed code verbatim, introducing license compliance issues into the project. At least two respondents felt that the models should be responsible for avoiding such problems. 
\resp{665} noted, ``Developers should not bear the burden for [...] copyright [...] concerns. Tools should not be producing code which could get a developer in trouble.''

Another avenue would be for developers to create or choose new OSS licenses that provide specific conditions for inclusion in training data. Such licenses would enable creators to control how and when their works are used for GenAI. As \resp{587} put it: ``I think that open source developers that hold strong opinions about the use of their code with AI models need to update their licenses to enforce those desires.'' \resp{429}, by contrast, conveyed that developers' preferences should be honored regardless of the existence of prohibitions in the license: ``[I]f an individual or institution wishes to have their data removed from a training dataset, they should have the right to request its removal.'' (The response did not suggest how such removal might be accomplished.) 
\looseness=-1

\finding{
\label{finding:attribution}Some respondents generally wanted licensing terms to be complied with by GenAI models, especially when producing code identical to or very similar to work in the training set.}

\subsubsection{Fair use and training data}
\label{subsec:fair_use}
While some developers, as described above, believed that the use of code as training data should comply with license requirements, others expressed in open-ended responses that they considered such use to be a fair use, even if it was not always clear whether respondents referred to this concept in its legal sense or in a more general sense. For example, \resp{173} said, ``I believe that AI training utilizing any published content is a form of fair use and should be encouraged rather than litigated. The usage is entirely transformative rather than derivative.'' \looseness=-1

Some respondents who believed that using content as training data was fair use compared the practice to the process through which humans produce new works based on the knowledge they have accumulated over time. These respondents believed that it should be assumed that publicly available code is free to use for training and that the burden should be on developers to indicate otherwise. For example, \resp{26} said, ``Creators need to make efforts to protect data that should not be used in AI training.'' 

In follow-up interviews, one respondent elaborated that they attempt to exercise such control by limiting the types of projects for which they use GenAI tools, thereby restricting the tools' access to their code. \iresp{118} specified that when choosing tools to use, ``I'm not going to use a tool at work that's going to expose my work to risk [...], whereas if it's just a personal project, I'm a lot less concerned. I want to make sure I'm not going to violate anything major or there's nothing truly weird like they're going to own my code.'' %

\finding{\label{finding:fair_use}Some developers believed that using code as training data should be considered fair use by default and that the onus should be on developers to protect code from being included in training data.}

\subsection{AI-generated output that resembles existing code}

Developers also had a range of views on whether it is wrong for generated code to closely resemble existing code. Some respondents compared what GenAI models do when producing code to what humans have always done (manually) in creating derivative work. \resp{103} stated: ``Being an open-source software developer, I'm accustomed to seeing my codes copied or `cloned' by other developers without any notice, and I don't feel anything special from it.%
'' \resp{265} noted, ``What AI is doing now was being done by developers [throughout] the years through forums, [S]tack [O]verflow, [G]ithub, etc.''  %
\looseness=-1

\finding{A group of developers did not see how GenAI technology was doing anything significantly different from traditional means of creating code, which reuses existing code created by others.} %

For other respondents, their view of the wrongfulness of generated code that resembled code they had developed depended on the nature of the code being copied, an intuition that tracks U.S. legal doctrine \cite{usc102b}. \resp{183} stated that ``small pieces of code are rarely valuable in themselves. For any given function, there's only so many ways to write it. It’s the entire thing that’s valuable.''  \resp{246} provided a concrete example, saying, ``There are only so many ways you can create a fetch request in Java[S]cript.''
\resp{573} summarized the concept: ``Development is typically a process of piecing together various snippets that are customized to a particular use case; there is no expectation of `ownership,' at least in the case of partial code blocks. While the reproduction of an entire application/code base for use in generating revenue is problematic, the sharing of common code is how the development community has progressed [until] now.'' And \resp{445} offered a useful analogy: ``[A]dding a license to a particular piece of code does not make sense. You can add licenses to libraries and exact forms of algorithms. Adding a license to a piece of code is like adding a copyright to individual strokes and colors in a painting.''

For some, the extent of copying was important, perhaps reflecting U.S. copyright law's concept of \textit{de minimis} copying, which does not constitute infringement. \resp{144} put it this way: ``I feel like if significant pieces of code are copied, only then does it really become an issue for me. I would not be comfortable if someone took multiple files from my code and used them in a proprietary project.'' \resp{143} said, ``If a generative AI happens to have copied large swathes of code, yes [the developers of that code deserve attribution].''

\finding{Some developers' views on the copying of code track U.S. copyright law: basic functions and building blocks are not protectable and should be free for all to use.}

\begin{figure}[t]
\centering
\includegraphics[width=\linewidth]{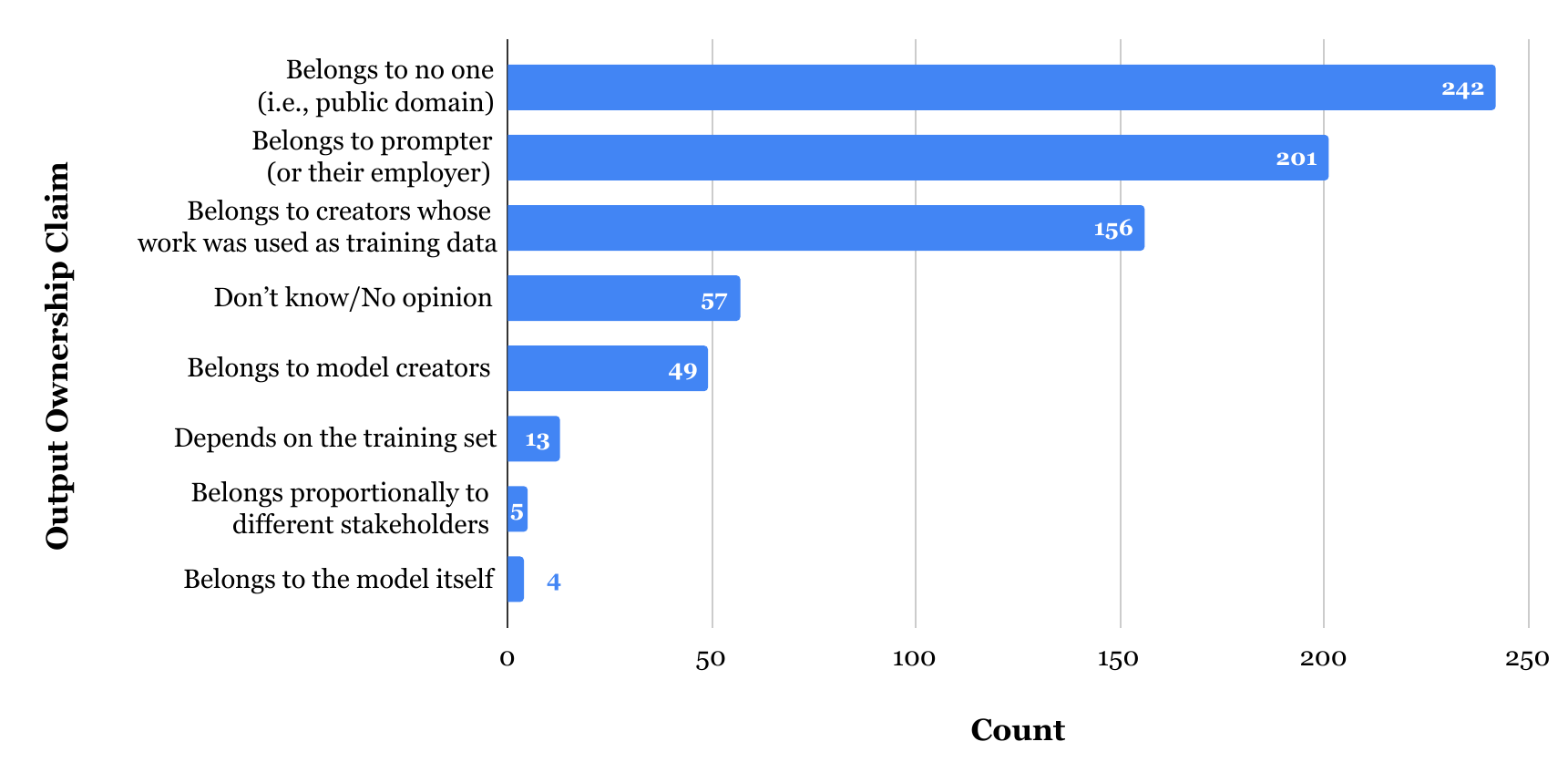}
\caption{Developer perceptions on output ownership}%
\label{fig:ownership}
\end{figure}

\subsection{On the ownership of AI-generated code}

Our discussion to this point has focused on developers' views on potential liability for copying another's code, either in training data or in the generated output.  Another important legal question is who would own any copyright to code created by GenAI models. In this subsection, we do not attempt to provide a legal analysis of this question; instead, our goal is to understand what developers think the answer to the question should be.

Participants were presented with a list of potential options as to who owns the copyright to code generated through the use of AI and were invited to select all that they believed applied, with the opportunity to provide other options and to further explain their answers.  Although courts have yet to resolve this question, only 57 of 554 respondents familiar with GenAI tools (10\%) indicated that they were not sure who owned the output or had no opinion on the matter. The rest of the respondents expressed an opinion and reported high confidence in their answers: 416 (75\%) stated that they were confident or even very confident.  \Cref{fig:ownership} shows respondent views on the ownership of model output.

\subsubsection{No one---works are in the public domain}
The most popular choice, selected by 242 participants~(44\%), was that AI-generated code belongs to no one. 
\resp{516} offered a summary: ``Code generated through the use of AI is not copyrightable, because it fails the creative step required by law of being a product of the intellect. Instead, it is produced by a non-sentient machine, lacks a creative step, an expression of the soul. It is therefore in the public domain.''

Others, like \resp{140}, selected this option to represent that they viewed the question as minimally important: ``It's [going to] be ridiculously difficult to prove anything, and humanity's efforts should be spent on advancing technology and society, not fighting silly legal battles about who gets to call dibs on random binary stuff.''

Several respondents invoked the inherently collaborative process of generating code through AI as a reason not to assign copyright ownership to only one individual or entity. \resp{74} gave an analogy: ``It'd be like saying who gets the copyright of what a child creates. [T]heir teachers? [T]heir parents? [T]heir friends? [T]hey all influence it in an unpredictable way. At least to my knowledge[,] we can't measure it.'' Putting it in more concrete terms, \resp{105} stated that the output of GenAI models is ``a product of complex interactions between AI algorithms, training data, and user inputs, making traditional copyright assignment impractical. This perspective aligns with the open source ethos, favoring communal access and innovation over individual ownership in the rapidly evolving field of AI and software development.'' 

This sentiment regarding the nature of collaborative work was also echoed in responses, particularly from open-source developers, that seemed to suggest that even if developers could claim copyright in AI output, they should refrain from asserting their rights in favor of the greater good. %
\resp{233} asserted that the focus should be maximizing utility: ``I've been contributing to open-source code for over a decade, consistently opting for the least restrictive licenses. I firmly believe in the value of sharing knowledge rather than imposing limitations. With the emergence of AI as a lasting tool, my stance is to focus on maximizing its utility for people rather than getting bogged down in minor details.''
\looseness=-1

\finding{The largest group of developers (242, 44\%) thought that the output of GenAI belongs in the public domain.  Some align with this view because it is consistent with the collaborative nature of open-source software development, while others believe that because AI models are not human, the output of those models cannot be protected by copyright law.}

\subsubsection{The prompting developer or their employer}
\label{subsec:prompter_owns}

The next most popular suggestion was that the model output belonged to the developer that prompted the model or their employer, as selected by 201 respondents (36\%). 

Many participants who reached this conclusion believed that using code-generation tools is similar to existing processes that can result in copyright-compatible code.  Respondents with this perspective, like \resp{173}, characterized GenAI as ``a tool like any other, used to augment the skills and direction of its user,'' where the assumption would be that the user owned the rights to whatever was created using the tool.  \resp{22} provided an analogy: ``[I]f a carpenter chooses to use a powerful automation tool like [a] CNC machine to make something, the final product is his (or his employer's), not the maker of the tool or [whoever] inspired the carpenter.''  Ultimately, \resp{118} indicated, ``If I can copyright code I learned by essentially copying what I learned from Stack Overflow, I should be able to copyright what I get from [...] Copilot. [...] If I can't, then the analysis should be similar. The focus on AI as the issue seems misguided.  The only difference is scale.''

Six developers explicitly stated that the prompt was an important starting point for the ownership question.  \resp{187} explained, ``[C]ode generation requires prompt work, and the copyright on the code that is generated should be the result of that prompt work.'' For other developers, the claim to ownership relied on the assumption that code resulting from GenAI is not ready to use immediately upon generation but must be further refined by the developer. As explained by \resp{107}, ``AI code generation typically doesn't result in a perfectly working code. It's a sketch, or draft variant that will be modified and fine[-]tuned [...]. 
AI doesn't do development on its own[;] it helps developers do their job.'' \resp{63} concluded that if the models were capable of ``creating a large amount of code without [requiring] editing,'' copyright issues might be more complex, but \resp{63} was ``not convinced that AI code generation is good enough for that.'' \resp{39} stated similarly, ``The development process of using AI development tools usually involves AI generating an incorrect outline of one snippet of code in a larger system that I then amend, correct and test. The additional work I put into it seems like it should constitute a new creative work that is copyrightable.''
\looseness=-1

\finding{%
\label{finding:copyright_user}
The second largest group of developers (201, 36\%) believed that the user of a GenAI tool should own the copyright to code generated with that tool because it is the user's contributions---creating the prompt and, typically, editing and further refining the result---that create the code. In other words, GenAI was seen as simply a new tool for creating content.}

One challenge raised by respondents, however, is that even if those supplying the prompts are deemed to be the owners of copyright in the resulting code, the nature of code creation with GenAI is such that similar code might result from two independently created prompts. Such activities could result, under U.S. law, in two separate copyrights, one for each generated work. \resp{98} found this legal conclusion to be counterintuitive: ``[T]he developers who prompted can't also be the copyright owners---it's fairly simple to get very similar or identical outputs from [a] LLM by two different developers.'' \resp{507} wrote, ``If I prompt the model to write code that I later sell, can I copyright it? Yes of course. [...] Can someone else prompt it the same way and maybe get the exact same code? Yes. Does that make sense? No. Which makes me think that AI copyright makes no sense.''

Ultimately, some respondents noted that control over the resulting work incentivizes creative activity. \resp{300} asserted, ``AI will not be a useful product to sell if the authors of the original training data or the model creators retain copyright on generated stuff, or if the output is uncopyrightable.''  \resp{114} mentioned that they ``specifically use [c]ode [g]eneration tools that include in their ToS specific statements that copyright belongs to us (the customer),'' and \resp{288} stated, ``If the license does not end up with me at the end of the circus, I will use a different tool made by a different company trained off different data.''
\looseness=-1

\subsubsection{The creators whose works make up the training data}

The next largest group, 156 respondents~(28\%), believed that rights in the generated code should belong to the individuals who created the code on which the GenAI model was trained. A common theme among such responses was a sense of fairness or justice, given GenAI's reliance on training data. As \resp{77} put it, ``[I]f the code is extremely similar or identical to the training data[,] then the copyright should belong to the respective creator of this training data.'' \resp{13} wrote, ``[I]f these humans had not written this code, this model would not exist.''

Some respondents, on the contrary, invoked the collaborative nature of development as a reason to deny copyright ownership claims by developers of the code used in the training data. \resp{157} noted, ``[T]he neural networks require so much input for them to work that none of the input is really deterministic to the end work. [...] So, the copyright cannot belong to the people [who] just provided the training data. [If that were true], the copyright of [a] book could belong also to the authors of all the books that the author had read before.''  \resp{157} continued by providing another analogy from the creative arts: ``[We don't give] copyright to the people that cleaned the theater and people that create the music[al] instruments and so on.''  

Respondents also noted that it may be difficult, if not impossible, to extract an exact mapping from the training examples to the generated output, particularly in situations with many similar or identical code instances (such as GitHub forks). As \resp{502} asked, ``[H]ow can we be so sure whether that code was originally written by that user?''

\finding{Some developers believed that fairness requires giving ownership of generated code to those developers who created the training data. However, others invoked the collaborative nature of development to oppose the idea of copyright claims by those whose code was used in the training data and highlighted the practical problems with mapping the training data to generated code to determine ownership.}

\subsubsection{The creators of the model}
The least popular of the provided options, ranked even below the ``don't know/no opinion'' option, was that the creators of GenAI models should own the output, with only 49 participants (9\%) indicating that they believed this to be the case. 
\resp{441} believed that ``the creators of the model should receive monetary compensation for training expenses.''
\looseness=-1

By contrast, other respondents, such as \resp{306}, indicated that while the ownership question is complicated, the creators of the models were the only group with no real claim: ``Again, it is difficult to be sure. What I do believe, however, is that it should never belong to the creators of the model, as they have no say on the output of the model, only its inner workings.'' \resp{98} expressed a similar sentiment: ``The creators of the model are definitely not the copyright owners---LLMs are just text generators and as such can technically generate any text. You can force [a] LLM to `write' literally any sequence of words you want---that [output] can't be copyrighted [by the model].'' 

\resp{159} offered a pragmatic view: ``If the code always belongs to the company that created and trained the model, there's no point in any company using it, especially freelance programmers and individuals.'' And \resp{157} suggested that if AI model creators held the copyright to generated code, that would mean for ``any picture created by a camera, the copyright would also be held by the camera manufactur[er]. And we do not do this.''

\finding{The model vendors who develop GenAI models were the one group that the fewest %
participants (49) believed had an ownership claim to generated outputs.}

\subsubsection{Other ownership claims}
Participants who selected multiple options conveyed that the answer to this question is often case-specific. Previous work exploring how changes in AI supply chains can influence the answers to copyright questions \cite{lee2023talkin} supports this view. \resp{182} provided scenarios in which the ownership of generated code might vary: ``Code generated by a model that I've prompted with a simple prompt (\eg `Please show me a solution to the Fizz-Buzz challenge') should probably be public domain. However, code generated based on a prompt that includes significant copyright content (\eg `Consider the style and techniques in the 50kB source file below, and then using those styles and relevant techniques, implement a new routine to [do something]. Before you start ask me any clarifying questions you think are important. [...]') could reasonably be protected as if it had been written by a human.'' \resp{152} wrote, ``[It] depends on the jurisdiction, and on how much of the code was generated by the AI, and on how much the output was transformed by the prompter.''
\looseness=-1

Some respondents chose multiple options because they believed that copyright ownership should be distributed among multiple groups. For example, \resp{191} called generated code ``a co[-]creation where we would have a percentage of ownership to apply. [...] The prompt, however, [would] belong entirely to [the] user.'' 
\resp{527} suggested, ``I would opt [for] [...] a shared benefits model, like [with] royalties in the entertainment industry, where the actual code owner gets a fair share, the model owner, and the end user who was capable of adding the right prompt into the model [...].'' Such views were uncommon, and no respondent proposed a system for determining an appropriate allocation of ownership rights or the implications thereof.

\finding{Some respondents believed that ownership depended on various factors, including the relevant jurisdiction, the amount of code generated, and how much the prompting developer contributed to the result. Others believed that multiple parties should share ownership claims and that a royalty-style system should be put in place.}

In the end, as \resp{54} noted, as GenAI becomes an increasingly common feature of all software, the question of who owns GenAI output may cease to matter: ``[T]he issue [of ownership] becomes more clouded, though, when you consider the current situation, where most generative AI is done through a service[, n]ot directly with the model. Therefore, the service that is providing the generated code or response can define in their own terms [of service] whether they own [the output] or not.''

\subsection{On the ownership and copying of prompts}
Much of the current GenAI litigation, as of this writing, is related to the unauthorized use of preexisting work in training data or the copyright status of generated content. However, we were also interested in understanding developers' views on the ownership of prompts, which, depending on their complexity, could be protected by copyright law. (As noted above, however, platforms' Terms of Service might govern users' rights in this regard.) Approximately half of the 554 respondents familiar with GenAI tools (278, 50\%) were indifferent as to whether a prompt they developed could be used by someone else without their permission, with 88 respondents (16\%) indicating that they would be pleased if such use occurred and 63 (11\%) extremely pleased. An additional 89 respondents (16\%) would be displeased by the use of their prompts, and 36 (7\%) would be extremely displeased. In general, as shown in \Cref{fig:likert_2}, most of the developers in our survey stated that they would not care (or even be pleased) if their prompts were reused by others, even without permission. %

\finding{Most respondents (77\%) indicated that they would be indifferent or even pleased if their prompts were copied and used by others, even without their permission.}

\subsection{Perceived risk of potential copyright infringement}%
In our follow-up interviews, five out of seven respondents perceived that, ultimately, the likelihood of legal enforcement resulting from the improper use of GenAI was low. %
\iresp{39} stated in an interview: ``If I technically am violating somebody's copyright for code that is never going to be pushed anywhere public, it's very implausible that I would be sued for that.''  \iresp{31} noted that they belonged to ``a fairly small organization. We're not high on the radar. I don't think that we're at any legal risk of anything. If anybody's going to get sued, it's not going to be us. It's going to be other people.'' \iresp{149} expressed a similar view: ``[T]he likelihood of detection is so small that it's no longer a key concern.'' 
\looseness=-1

\finding{Individuals and smaller organizations may perceive limited practical legal risk from using AI-generated code due to the low likelihood that any code that potentially infringes on another's copyright or violates another's license would be discovered.}

\section{\ref{rq:3}: Other legal concerns that developers anticipate as the use of generative AI increases}
\label{sec:rq3}

The legal issues other than those related to copyright law that GenAI raises are apparent. \ref{rq:3} aimed to discover which of these concerns were top of mind among the developers in our study. %
Developers identified a wide range of issues, including those relating to tort liability, malicious use, and data privacy concerns. Findings 25-29 answer this research question, a summary of which is also available in our replication package \cite{anonymous_repo}.%

\subsection{Liability concerns}
Respondents mentioned concerns about possible tort liability (legal responsibility for harmful actions) related to the use of GenAI. Product liability (the claim, for example, that a product is defective or harmful) was mentioned by 14 respondents (3\%), and liability for code generated with bugs or defects was mentioned by 13 respondents (2\%). In particular, developers were concerned about who would be held responsible for security issues or other bugs introduced by AI-generated code. \resp{0} expressed that they ``could see in the near future running [into] issues with security breaches caused by an AI code generator making suboptimal choices. A contrived example[,] but imagine [that] someone uses the AI to build a log-in. The AI suggests that you store passwords in plain text in your database which is later leaked. The developer could argue [that] the model creator (e.g., [O]penAI) was responsible for the breach.''

Other respondents identified potential issues with AI agents, such as consumer-facing chatbots that appear to engage in contractual agreements with users \cite{chatbotRefund, chatbotProblems} or AI agents that appear to operate autonomously, such that parties may attempt to disclaim responsibility for harmful effects. \resp{149} said that ``[f]unction calling''---using LLMs ``to write and call functions''---was a potential concern (see examples in \cite{autogpt, gptengineer}) and asked, ``[W]hat happens if those called (which may have real world effects) cause harm---who is responsible?''  Respondents also highlighted the risk posed by using GenAI in safety/mission-critical systems and domains, such as aviation, security, defense, and medical fields.

Finally, 13 respondents (2\%) were concerned that code might be generated without sufficient consideration of compliance issues, such as ``ignoring GDPR [the EU's General Data Protection Regulation]'' (\resp{190}).
\resp{665} was concerned that ``[i]f an AI generates specifications for a project, it may not consider existing laws and [so] generate software which does not comply with regulations.''  
Several respondents raised the related concern of using GenAI to draft legal documents such as software licenses, terms of service, and privacy policies, which could not only lead to an undesirable proliferation of such documents but may also, if not subject to sufficient human review, result in language that does not have the desired legal effect.

\finding{\label{finding:liability}Some developers raised questions about who would be held accountable for bugs and other defects in GenAI outputs, especially for mission-critical systems such as medical, aviation, and security. Some developers were also concerned about who would be held responsible for actions that AI agents take autonomously.}

\subsection{Generation of malicious content}  
Seventeen survey respondents also highlighted the potential that GenAI tools would be used to produce a variety of malicious content, including malware or viruses, disinformation, illicit or unsavory material, and propaganda.  While not a novel insight~\cite{fritsch2022overview,goldstein2024persuasive,pa2023attacker}, developers are aware that pre-existing technical and skill barriers to creating malicious content have been lowered, enabling less experienced individuals to carry out more sophisticated attacks with little to no domain knowledge. %
While many mainstream LLMs have guardrails in place to prevent the generation of this type of content \cite{openai_guardrails}, unrestricted, open-weight models such as the Dolphin family of models \cite{dolphin_uncensored} and alternatives such as WormGPT \cite{wormgpt} can still be used by malicious actors, and jailbreaking techniques can be used to bypass safety rails even for more widely used models~\cite{wei2024jailbroken, liu2024hitchhiker}.  

\finding{Some respondents identified the purposeful generation of malicious content as a societal problem and legal concern going forward.}

\subsection{Data privacy, information leakage, and discovery of system vulnerabilities}%
Information leaks were mentioned by 33 respondents (6\%), and data privacy concerns by an additional 15 respondents (3\%). This concern has been explored in the existing literature \cite{Abascal2023TMIFM,carlini2021extracting}, but it is of note that it is also a real-world consideration for developers when choosing which GenAI tools to trust and use.  Possible issues include proprietary code or information being produced by GenAI, leaked secrets such as API tokens and passwords, and the release of personally identifiable information. This risk results, as \resp{447} put it, from the fact that the ``devs [...] didn't realize [sensitive information would] get hoovered up'' in the training data. 

\finding{\label{finding:information_leakage}Developers and other stakeholders may not intend or anticipate that their data would be collected \emph{en masse}, so some developers are concerned that private and sensitive information can easily make its way into training sets.}

Similarly, there were concerns that generative models could be used by malicious actors for hacking (6), reverse engineering (6), and jailbreaking (1) systems. \resp{393} said that people could use ``AI to hack/discover security vulnerabilities [...].'' %

\finding{Generative AI, particularly LLMs, can not only leak information that was contained in the original training data but can also be used by malicious actors to exploit systems.}

Respondents were also cognizant of the dangers of including private or proprietary information in prompts, fearing that those same prompts would be used in training sets of future models. \iresp{31} explained, ``Because I'm almost only doing work stuff, and in the enterprise environment, I can't run the risk of the training set pulling security keys and that kind of thing. Because it's been demonstrated that you can trick those things into outputting data from the training set.''  

\subsection{Testing and test generation}
Tasks related to testing, such as generating test cases, were identified by 28 respondents (5\%) as having potential legal ramifications. Some respondents identified software testing itself as a shortcoming of GenAI's capabilities; as \resp{478} put it, ``[W]ho will test the generated test?'' Beyond this, participants asserted that the application of GenAI for testing could lead to legal problems in several ways, including use for penetration testing (\resp{299}) and security audits~(\resp{364}). The specter of liability also emerged (\resp{314}): ``AI models can automate testing processes, but if automated testing leads to false negatives or positives that result in financial or reputational harm, it may result in legal disputes.'' %

\finding{Some developers had concerns related to AI-generated tests, including their completeness and reliability.}

\section{Threats to Validity}
\label{sec:threats}

\subsection{Construct validity} 
Given the design of our study, the results represent developers' communicated perceptions, which may or may not reflect their or their organizations' actual practices. We have avoided making any claim connecting such perceptions to actual practices. While we sought to include a wide range of perspectives, the demographic information included in this study was self-reported by respondents and was not independently verified. Absent a foolproof method of prohibiting duplicate responses, it may have been possible for a participant to submit more than one response. However, no incentive was associated with the survey that would motivate such behavior, and, to the best of our knowledge, we received no duplicate responses. We also acknowledge that neither the survey respondents nor the interviewees may constitute a representative sample of developers. To mitigate this threat, through our coding process, we aimed to report those views and perceptions that were held by several developers. Lastly, our respondent sample may overrepresent developers who frequently use GenAI tools and underrepresent those who rarely or never use GenAI tools.%
\looseness=-1

\subsection{Internal validity}
We followed best practices when developing the questionnaire to ensure that the questions were straightforward and would not bias the participants toward a particular answer. We also conducted a small pilot study with graduate students in our research lab to check readability, length, and content. 
Given that participation in the survey was on an entirely voluntary basis and derived from a single source of possible respondents, we are aware of the problems introduced by self-selection bias. However, the number of participants gives us confidence that the results capture various developers' views. In the qualitative analysis of survey responses, we mitigated confirmation bias by having more than one annotator examine the data independently and then reconvene to discuss and resolve any disagreements, rigorously following the best practices of qualitative analysis. Our conclusions were derived from data analysis.
\looseness=-1

\subsection{External validity}
To investigate developers' perspectives, we rely on GitHub developers who showed interest in repositories of GenAI tools for coding. While our respondents may represent developers working in different domains and organizations, we are also aware that they may overrepresent views typical of open-source ecosystems and underrepresent those of closed-source developers. Although we do not claim generalizability,  many respondents reported working for companies or in education, which may lead to perspectives that vary from those in the open-source community more generally. Moreover, while we had participants who reported being from 73 countries spanning six continents%
, nearly a third of respondents came from the U.S., so our results may apply most directly to the U.S. context and legislation.

\section{Discussion and Implications}
\label{sec:implications}

We distill the key issues highlighted by developers, as reported in our study, and discuss their implications. We also examine potential directions for future research in light of the identified challenges.
\looseness=-1

\subsection{Developing new licenses for software used in GenAI training}
Because open-source software (OSS) is used to train GenAI models, licensing is an important consideration. Respondents expressed a range of views on the permissibility of using OSS as training data 
(Finding \ref{finding:training_data}),
with some condoning such use (Finding \ref{finding:fair_use}) and others believing that OSS developers should take proactive steps if they want their work to be excluded from such use, such as including restrictions in a license (Findings \ref{finding:monetary_compensation} and \ref{finding:attribution}). For many years, there has been a standard set of licenses in the OSS community, and drafting new licenses (``license proliferation'') has been discouraged, as it makes compliance tasks more difficult~\cite{prevalent,wintersgill2024law}. However, the text of the most popular OSS licenses was drafted decades ago and did not anticipate technological developments such as GenAI. This means that it may be unclear whether or how the terms of existing OSS licenses apply to scenarios in which OSS projects are used to train, develop, and distribute AI systems.   

In the past, updated versions of popular licenses have been drafted to account for new technology~\cite{whyagpl,whygplv3}, and AI model-specific licenses (\eg OpenRAIL~\cite{openrail}) have emerged that impose ethical usage requirements. Thus, the new legal challenges introduced by large-scale training and GenAI may necessitate the drafting of new OSS licenses or the revision of existing licenses so as to provide guidance on these issues and allow developers to more clearly indicate their views on the use of their work in connection with GenAI. Importantly, since some existing OSS licenses were drafted from the viewpoint of the OSS community, their terms are sometimes difficult to align with legal doctrine, leading to a perceived need for drafting organizations to issue FAQs explaining the intent of their licenses~\cite{wintersgill2024law}. If those with legal expertise, preferably with accompanying SE backgrounds, participate in drafting these new licenses, it is possible that ambiguous and unclear language can be avoided to the largest extent possible.

\subsection{Disclosing training data}
\looseness=-1
Participants who had a view about the use of their code in training data communicated that it was important to respect the terms of the license, including the attribution requirements. 
(Findings \ref{finding:monetary_compensation}, \ref{finding:attribution}, and \ref{finding:fair_use}).
Absent further recognition of attribution rights under U.S. copyright law, licensing will continue to be a key method for creators to enforce this preference. However, for this to occur, the content and provenance of training data must be made available to the users of the models, including information about the source of each piece of training data, the license or licenses under which the data were originally issued, and the historical record of the data throughout various processes.  

Although regulators should consider the importance of attribution to creators, they should also recognize that any legal requirement to provide complete information on the content in training material will involve significant technical hurdles. (In fact, as explained in \Cref{sec:legal_background}, the EU AI act calls for the transparency of models and, therefore, recommends that training datasets be disclosed.) DataBOMs (Data Bills of Materials)~\cite{stalnaker2024boms,barclay2019towards} are a conceptual solution but cannot easily be deployed at the scale required to capture the billions, if not trillions, of data points on which LLMs are trained ~\cite{devilInDetails}. Additionally, any such provenance information must take into account that a work might appear in different ways: on a site created by the developer, in a StackOverflow post authored by another user, or as part of another developer's work. A single DataBOM is not likely to capture these various uses.

Finally, some developers may favor the recognition of a right to be removed from training data (as foreseen by legislation such as the EU AI Act). Apart from the difficulty of determining whether a work is included in a training dataset, the further questions of what it means to be ``removed'' and how such removal can be effectively accomplished remain unanswered (and perhaps unanswerable, absent a complete destruction of the training data) ~\cite{lee2023talkin,yao2023}. This reality may align with some respondents' views that the use of this technology may outpace at least some attempts to regulate it.
\looseness=-1

\subsection{Tracking data provenance}
As described above, tracking GenAI's provenance of data output is technologically challenging. Although such functionality has been achieved by newer models, such as ChatGPT-4 and Gemini, this is based on retrieval-augmented generation (RAG)--style systems \cite{ma2024crafting}, which essentially depend on a database or web search to provide additional context to the model. Models employing RAG extract and/or summarize content obtained from these searches and then provide provenance information for the search results used to generate the final output. This is different from providing provenance information for data points that were included in the training data. %

The process of retrieving provenance for model generations that do not rely on external data sources accessible through RAG or provided in a prompt is still an open research area. As such, absent external, supplied provenance information, models may not be able to cite a specific source from which their output was derived. However, such information could be useful to address several concerns raised by developers, including being notified of information leakage (Finding \ref{finding:information_leakage}) or determining the root of bugs or vulnerabilities (Finding \ref{finding:liability}).

Since new LLM models can be created by fine-tuning existing base models, it also becomes necessary to track the provenance and lineage of models appropriately. Failure to do so could result in not all training data being known or disclosed. These relationships emerging from the AI supply chain could be tracked using standardized model cards \cite{mitchell2019model} or AIBOMs (AI Bills of Materials)~\cite{xia2023empirical, stalnaker2024boms}, although these methods add their own compliance challenges.

Tracking the provenance of generated code requires knowledge of the presence of generated code. However, developers reported that the use of GenAI to produce code is rarely documented (Finding \ref{finding:no_documentation}), and even in cases where such documentation exists, it often notes only that GenAI was used without providing details of the prompt or the rest of the interaction (Finding \ref{finding:documentation_process}). This could obscure the presence of generated code, complicating provenance and making it more difficult to achieve developers' goals surrounding attribution, license compliance, and risk mitigation concerning generated code. Future work investigating the process by which GenAI is utilized in software engineering may help to identify methods by which to facilitate documentation of GenAI usage that are useful for provenance efforts.

\subsection{New AI advancements will likely amplify legal challenges}
Many of the study participants expressed the sentiment that ``tools are tools'' and, as such, developers should hold the copyright to GenAI output (Finding \ref{finding:copyright_user}). Much of this understanding is based on the current limitations of existing tools, which, in many cases, require developers to manually correct, edit, or adapt the solution provided by AI before the code can be used. However, some respondents indicated that they anticipated more complicated legal challenges ahead once technology advances and can produce entire software packages composed of multiple files that closely resemble existing licensed components, a process already under development~\cite{devin, gptpilot}. The question will then be whether developers' attitudes about generated code ownership will change based on the nature and scope of developers' own contributions. Likewise, while developers' intuitions about code usage may currently align, at least to some extent, with U.S. copyright law, there may well be a future misalignment as both the technology and the law evolve. Finally, our study indicates that incentives relating to creativity and the use of preexisting work might vary considerably depending on the community's norms, such as whether the work is open source or proprietary. Having a better understanding of the expectations and perceptions of developers could help guide regulators and decision-makers in this area and raise questions about whether ``one size fits all'' is the right approach. 

Several study participants also emphasized the speed with which AI development is taking place and whether law and regulation would be able to keep up (Finding \ref{finding:litigation_and_speed}). Other respondents expressed the view that GenAI was different only in degree and not in kind from other creative processes that do not attract regulation. Still others reported confidence in their views on as-yet unresolved legal issues regarding copyright and GenAI. Collectively, such respondents seemed to suggest that the law as written may not be a strong motivator in this development space. Decision-makers might therefore consider whether and how regulation can address concerns without being seen as an obstacle merely to be designed around.

Finally, to the extent that copyright law is concerned with incentivizing creativity, GenAI regulators should not think about copyright law in a vacuum. The other legal concerns raised by developers---such as concerns about privacy, security, and tort liability (Finding \ref{finding:liability})---are likely to have an effect on how and when they use GenAI and may, in fact, work at cross-purposes with the motivations underlying copyright law. For example, developers might be conflicted about asserting authorship over GenAI work for copyright purposes if that means also bearing responsibility for unforeseen harms.

\subsection{Final thoughts and implications}
Although there are many open questions regarding the use of GenAI and copyright, the main priority for future research should be to address problems related to data provenance. Developers expect tools that can provide them with citations and will not lead them unwittingly into legal trouble. Likewise, they need tools that allow them to document and track the code generated by LLMs for transparency and control of legal risk. Given that this paper primarily examines these issues from a perspective grounded in U.S. law, we encourage further research to interpret our results against regulatory efforts elsewhere.
\looseness=-1

\section{Conclusion}
\label{sec:conclusion}

In this paper, we present a study, conducted by a joint team of SE and legal researchers, which surveyed 574 software developers from around the world who use GenAI tools for various software development tasks. Through an online survey and follow-up interviews, we explored developers' perspectives on emerging legal issues, including their perception of copyrightability, ownership of generated code, and other related concerns. Our aim was to identify potential misconceptions among developers, assess the impact of GenAI on their work, and evaluate their awareness of licensing and copyright risks. Using both qualitative and quantitative methods to analyze the responses, we found that opinions on copyright issues vary widely, especially regarding the ownership of model outputs. In addition, many developers demonstrated an awareness of the complexities and nuances involved in addressing these increasingly complex legal questions. We provide a discussion of our findings that can inform policy decisions and guide future research.

\section{Acknowledgments}
We would like to thank the study participants for their time and valuable contributions. This research has been supported in part by NSF grant CCF-2217733. Any opinions, findings and conclusions expressed herein are the authors and do not necessarily reflect those of the sponsors. A complete and detailed list of image attributions can be found in our online replication package~\cite{anonymous_repo}.%

\bibliographystyle{ACM-Reference-Format}
\bibliography{references}


\begin{thebibliography}{132}

%%% the Web address.

\ifx \showCODEN    \undefined \def \showCODEN     #1{\unskip}     \fi
\ifx \showDOI      \undefined \def \showDOI       #1{#1}\fi
\ifx \showISBNx    \undefined \def \showISBNx     #1{\unskip}     \fi
\ifx \showISBNxiii \undefined \def \showISBNxiii  #1{\unskip}     \fi
\ifx \showISSN     \undefined \def \showISSN      #1{\unskip}     \fi
\ifx \showLCCN     \undefined \def \showLCCN      #1{\unskip}     \fi
\ifx \shownote     \undefined \def \shownote      #1{#1}          \fi
\ifx \showarticletitle \undefined \def \showarticletitle #1{#1}   \fi
\ifx \showURL      \undefined \def \showURL       {\relax}        \fi
% invisible to TeX
\providecommand\bibfield[2]{#2}
\providecommand\bibinfo[2]{#2}
\providecommand\natexlab[1]{#1}
\providecommand\showeprint[2][]{arXiv:#2}

\bibitem[cop(2023)]%
        {copyrightRegistration}
 \bibinfo{year}{2023}\natexlab{}.
\newblock \bibinfo{howpublished}{Letter from Robert J. Kasunic, Associate Register of Copyrights and Director of Registration Policy and Practice, U.S. Copyright Office, to Van Lindberg, Taylor English Duma LLP (February 21, 2023), \url{https://www.copyright.gov/docs/zarya-of-the-dawn.pdf}}.
\newblock


\bibitem[get(2023)]%
        {gettyVsStability23}
 \bibinfo{year}{2023}\natexlab{}.
\newblock \bibinfo{title}{{Getty Images (US) Inc. v Stability AI Inc}}.
\newblock \bibinfo{howpublished}{1:23-cv-00135 (D. Del.)}.
\newblock


\bibitem[ano(2025)]%
        {anonymous_repo}
 \bibinfo{year}{2025}\natexlab{}.
\newblock \bibinfo{title}{Online replication package}.
\newblock \bibinfo{howpublished}{\url{https://archive.softwareheritage.org/browse/origin/directory/?origin_url=https://github.com/TStalnaker44/copyright_and_genai_tosem_25}}.
\newblock


\bibitem[qua(nd)]%
        {qualtrics}
 \bibinfo{year}{[n.d.]}\natexlab{}.
\newblock \bibinfo{title}{Qualtrics}.
\newblock \bibinfo{howpublished}{\url{https://www.qualtrics.com/}}.
\newblock
\newblock
\shownote{Accessed: 2023-21-06}.


\bibitem[17 U.S.C. § 102(a)({[n.\,d.]})]%
        {usc102a}
17 U.S.C. § 102(a) \bibinfo{year}{[n.\,d.]}\natexlab{}.
\newblock \bibinfo{howpublished}{17 U.S.C. § 102(a)}.
\newblock


\bibitem[17 U.S.C. § 106({[n.\,d.]})]%
        {usc106}
17 U.S.C. § 106 \bibinfo{year}{[n.\,d.]}\natexlab{}.
\newblock \bibinfo{howpublished}{17 U.S.C. § 106}.
\newblock


\bibitem[17 U.S.C. § 1202(b)({[n.\,d.]})]%
        {usc1202b}
17 U.S.C. § 1202(b) \bibinfo{year}{[n.\,d.]}\natexlab{}.
\newblock \bibinfo{howpublished}{17 U.S.C. § 1202(b)}.
\newblock


\bibitem[17 U.S.C. §102(b)({[n.\,d.]})]%
        {usc102b}
17 U.S.C. §102(b) \bibinfo{year}{[n.\,d.]}\natexlab{}.
\newblock \bibinfo{howpublished}{17 U.S.C. §102(b)}.
\newblock


\bibitem[Abascal et~al\mbox{.}(2023)]%
        {Abascal2023TMIFM}
\bibfield{author}{\bibinfo{person}{John Abascal}, \bibinfo{person}{Stanley Wu}, \bibinfo{person}{Alina Oprea}, {and} \bibinfo{person}{Jonathan Ullman}.} \bibinfo{year}{2023}\natexlab{}.
\newblock \showarticletitle{TMI! Finetuned Models Leak Private Information from their Pretraining Data}.
\newblock \bibinfo{journal}{\emph{Proc. Priv. Enhancing Technol.}}  \bibinfo{volume}{2024} (\bibinfo{year}{2023}), \bibinfo{pages}{202--223}.
\newblock
\urldef\tempurl%
\url{https://api.semanticscholar.org/CorpusID:259064156}
\showURL{%
\tempurl}


\bibitem[Andersen v. Stability AI Ltd.({[n.\,d.]})]%
        {andersenVsStability22}
Andersen v. Stability AI Ltd. \bibinfo{year}{[n.\,d.]}\natexlab{}.
\newblock \bibinfo{title}{{Andersen v. Stability AI Ltd.}}
\newblock \bibinfo{howpublished}{3:23-cv-00201, (N.D. Cal.)}.
\newblock


\bibitem[Anthropic Models({[n.\,d.]})]%
        {anthropic_models}
Anthropic Models \bibinfo{year}{[n.\,d.]}\natexlab{}.
\newblock \bibinfo{title}{Models}.
\newblock \bibinfo{howpublished}{\url{https://docs.anthropic.com/en/docs/about-claude/models}}.
\newblock
\newblock
\shownote{Accessed: 2024-17-10}.


\bibitem[Authors Guild, Inc. v. Google, Inc.(2015)]%
        {2015authors}
Authors Guild, Inc. v. Google, Inc. \bibinfo{year}{2015}\natexlab{}.
\newblock \bibinfo{howpublished}{Authors Guild, Inc. v. Google, Inc., 804 F.3d 202 (2d Cir. 2015)}.
\newblock


\bibitem[Authors Guild v. OpenAI Inc.(2023)]%
        {authorsVsChatGPT23}
Authors Guild v. OpenAI Inc. \bibinfo{year}{2023}\natexlab{}.
\newblock \bibinfo{howpublished}{{Authors Guild v. OpenAI Inc., 1:23-cv-08292, (S.D.N.Y.)}}.
\newblock


\bibitem[AutoGPT({[n.\,d.]})]%
        {autogpt}
AutoGPT \bibinfo{year}{[n.\,d.]}\natexlab{}.
\newblock \bibinfo{title}{AutoGPT}.
\newblock \bibinfo{howpublished}{\url{https://github.com/Significant-Gravitas/AutoGPT}}.
\newblock


\bibitem[{B. Rozière et al.}(2024)]%
        {rozière2024code}
\bibfield{author}{\bibinfo{person}{{B. Rozière et al.}}} \bibinfo{year}{2024}\natexlab{}.
\newblock \bibinfo{title}{{Code Llama: Open Foundation Models for Code}}.
\newblock
\newblock
\showeprint[arxiv]{2308.12950}~[cs.CL]


\bibitem[Bandi(2019)]%
        {prevalent}
\bibfield{author}{\bibinfo{person}{Mahak Bandi}.} \bibinfo{year}{2019}\natexlab{}.
\newblock \bibinfo{title}{All About Open Source Licenses}.
\newblock \bibinfo{howpublished}{\url{https://fossa.com/blog/what-do-open-source-licenses-even-mean/}}.
\newblock
\newblock
\shownote{Accessed: 2023-24-09}.


\bibitem[Barclay et~al\mbox{.}(2019)]%
        {barclay2019towards}
\bibfield{author}{\bibinfo{person}{Iain Barclay}, \bibinfo{person}{Alun Preece}, \bibinfo{person}{Ian Taylor}, {and} \bibinfo{person}{Dinesh Verma}.} \bibinfo{year}{2019}\natexlab{}.
\newblock \showarticletitle{{Towards Traceability in Data Ecosystems Using a Bill of Materials Model}}.
\newblock \bibinfo{journal}{\emph{arXiv preprint arXiv:1904.04253}} (\bibinfo{year}{2019}).
\newblock


\bibitem[Barke et~al\mbox{.}(2023)]%
        {barke2023grounded}
\bibfield{author}{\bibinfo{person}{Shraddha Barke}, \bibinfo{person}{Michael~B James}, {and} \bibinfo{person}{Nadia Polikarpova}.} \bibinfo{year}{2023}\natexlab{}.
\newblock \showarticletitle{{Grounded Copilot: How Programmers Interact with Code-Generating Models}}.
\newblock \bibinfo{journal}{\emph{Proceedings of the ACM on Programming Languages}} \bibinfo{volume}{7}, \bibinfo{number}{OOPSLA1} (\bibinfo{year}{2023}), \bibinfo{pages}{85--111}.
\newblock


\bibitem[Belanger(2024)]%
        {chatbotRefund}
\bibfield{author}{\bibinfo{person}{Ashley Belanger}.} \bibinfo{year}{2024}\natexlab{}.
\newblock \bibinfo{title}{{Air Canada} Has to Honor a Refund Policy Its Chatbot Made Up}.
\newblock \bibinfo{howpublished}{\url{https://www.wired.com/story/air-canada-chatbot-refund-policy}}.
\newblock


\bibitem[Brittain(2023)]%
        {author_lawsuit}
\bibfield{author}{\bibinfo{person}{Blake Brittain}.} \bibinfo{year}{2023}\natexlab{}.
\newblock \bibinfo{title}{{Authors sue Meta, Microsoft, Bloomberg in latest AI copyright clash}}.
\newblock \bibinfo{howpublished}{\url{https://www.reuters.com/legal/litigation/authors-sue-meta-microsoft-bloomberg-latest-ai-copyright-clash-2023-10-18/}}.
\newblock


\bibitem[california-regulation(2024)]%
        {california_regulation}
california-regulation \bibinfo{year}{2024}\natexlab{}.
\newblock \bibinfo{title}{Governor Newsom announces new initiatives to advance safe and responsible AI, protect Californians}.
\newblock \bibinfo{howpublished}{\url{https://www.gov.ca.gov/2024/09/29/governor-newsom-announces-new-initiatives-to-advance-safe-and-responsible-ai-protect-californians/}}.
\newblock


\bibitem[Carlini et~al\mbox{.}(2021)]%
        {carlini2021extracting}
\bibfield{author}{\bibinfo{person}{Nicholas Carlini}, \bibinfo{person}{Florian Tramer}, \bibinfo{person}{Eric Wallace}, \bibinfo{person}{Matthew Jagielski}, \bibinfo{person}{Ariel Herbert-Voss}, \bibinfo{person}{Katherine Lee}, \bibinfo{person}{Adam Roberts}, \bibinfo{person}{Tom Brown}, \bibinfo{person}{Dawn Song}, \bibinfo{person}{Ulfar Erlingsson}, {et~al\mbox{.}}} \bibinfo{year}{2021}\natexlab{}.
\newblock \showarticletitle{Extracting Training Data from Large Language Models}. In \bibinfo{booktitle}{\emph{30th USENIX Security Symposium (USENIX Security 21)}}. \bibinfo{pages}{2633--2650}.
\newblock


\bibitem[codeium({[n.\,d.]})]%
        {codeium}
codeium \bibinfo{year}{[n.\,d.]}\natexlab{}.
\newblock \bibinfo{title}{codeium.vim}.
\newblock \bibinfo{howpublished}{\url{https://github.com/Exafunction/codeium.vim}}.
\newblock


\bibitem[cody({[n.\,d.]})]%
        {cody}
cody \bibinfo{year}{[n.\,d.]}\natexlab{}.
\newblock \bibinfo{title}{cody}.
\newblock \bibinfo{howpublished}{\url{https://github.com/sourcegraph/cody}}.
\newblock


\bibitem[colorado-bill(2024)]%
        {colorado_bill}
colorado-bill \bibinfo{year}{2024}\natexlab{}.
\newblock \bibinfo{title}{Consumer Protections for Artificial Intelligence}.
\newblock \bibinfo{howpublished}{\url{https://leg.colorado.gov/bills/sb24-205}}.
\newblock


\bibitem[Commission(2024)]%
        {eucopyrights}
\bibfield{author}{\bibinfo{person}{European Commission}.} \bibinfo{year}{2024}\natexlab{}.
\newblock \bibinfo{title}{{The {EU} copyright legislation}}.
\newblock
\newblock
\newblock
\shownote{https://digital-strategy.ec.europa.eu/en/policies/copyright-legislation}.


\bibitem[Constantino et~al\mbox{.}(2023)]%
        {constantino2023perceptions}
\bibfield{author}{\bibinfo{person}{Kattiana Constantino}, \bibinfo{person}{Mauricio Souza}, \bibinfo{person}{Shurui Zhou}, \bibinfo{person}{Eduardo Figueiredo}, {and} \bibinfo{person}{Christian K{\"a}stner}.} \bibinfo{year}{2023}\natexlab{}.
\newblock \showarticletitle{Perceptions of open-source software developers on collaborations: An interview and survey study}.
\newblock \bibinfo{journal}{\emph{Journal of Software: Evolution and Process}} \bibinfo{volume}{35}, \bibinfo{number}{5} (\bibinfo{year}{2023}), \bibinfo{pages}{e2393}.
\newblock


\bibitem[Cooper and Grimmelmann(2024)]%
        {cooper2024files}
\bibfield{author}{\bibinfo{person}{A.~Feder Cooper} {and} \bibinfo{person}{James Grimmelmann}.} \bibinfo{year}{2024}\natexlab{}.
\newblock \showarticletitle{The Files Are in the Computer: Copyright, Memorization, and Generative AI}.
\newblock \bibinfo{journal}{\emph{arXiv preprint arXiv:2404.12590}} (\bibinfo{year}{2024}).
\newblock


\bibitem[CopyrightCatcher({[n.\,d.]})]%
        {patronusCopyrightCatcher}
CopyrightCatcher \bibinfo{year}{[n.\,d.]}\natexlab{}.
\newblock \bibinfo{title}{{Introducing CopyrightCatcher, the first Copyright Detection API for LLMs}}.
\newblock
\newblock
\newblock
\shownote{Accessed: March 22, 2024. \url{https://www.patronus.ai/blog/introducing-copyright-catcher}}.


\bibitem[{Council of the European Union}(2024a)]%
        {aiact_article53}
\bibfield{author}{\bibinfo{person}{{Council of the European Union}}.} \bibinfo{year}{2024}\natexlab{a}.
\newblock \bibinfo{title}{{Article 53: Obligations for Providers of General-Purpose AI Models}}.
\newblock
\newblock
\newblock
\shownote{\url{https://artificialintelligenceact.eu/article/53/}}.


\bibitem[{Council of the European Union}(2024b)]%
        {aiact2024}
\bibfield{author}{\bibinfo{person}{{Council of the European Union}}.} \bibinfo{year}{2024}\natexlab{b}.
\newblock \bibinfo{title}{{Proposal for a Regulation of the European Parliament and of the Council laying down harmonised rules on artificial intelligence (Artificial Intelligence Act) and amending certain Union legislative acts}}.
\newblock
\newblock
\newblock
\shownote{\url{https://digital-strategy.ec.europa.eu/en/policies/regulatory-framework-ai}}.


\bibitem[cptX({[n.\,d.]})]%
        {cptX}
cptX \bibinfo{year}{[n.\,d.]}\natexlab{}.
\newblock \bibinfo{title}{cptX}.
\newblock \bibinfo{howpublished}{\url{https://github.com/maxim-saplin/cptX}}.
\newblock


\bibitem[Craig(2024)]%
        {craig2024ai}
\bibfield{author}{\bibinfo{person}{Carys~J Craig}.} \bibinfo{year}{2024}\natexlab{}.
\newblock \showarticletitle{THE AI-Copyright Trap}.
\newblock \bibinfo{journal}{\emph{Available at SSRN, \url{https://papers.ssrn.com/sol3/papers.cfm?abstract_id=4905118}}} (\bibinfo{year}{2024}).
\newblock


\bibitem[DOE 1 et al v. GitHub({[n.\,d.]})]%
        {doeVsGithub22}
DOE 1 et al v. GitHub \bibinfo{year}{[n.\,d.]}\natexlab{}.
\newblock \bibinfo{title}{{DOE 1 et al v. GitHub}}.
\newblock \bibinfo{howpublished}{4:22-cv-06823, (N.D. Cal.)}.
\newblock


\bibitem[Duan et~al\mbox{.}(2024)]%
        {duan2024membership}
\bibfield{author}{\bibinfo{person}{Michael Duan}, \bibinfo{person}{Anshuman Suri}, \bibinfo{person}{Niloofar Mireshghallah}, \bibinfo{person}{Sewon Min}, \bibinfo{person}{Weijia Shi}, \bibinfo{person}{Luke Zettlemoyer}, \bibinfo{person}{Yulia Tsvetkov}, \bibinfo{person}{Yejin Choi}, \bibinfo{person}{David Evans}, {and} \bibinfo{person}{Hannaneh Hajishirzi}.} \bibinfo{year}{2024}\natexlab{}.
\newblock \showarticletitle{{Do Membership Inference Attacks Work on Large Language Models?}}
\newblock \bibinfo{journal}{\emph{arXiv preprint arXiv:2402.07841}} (\bibinfo{year}{2024}).
\newblock


\bibitem[Ebert and Louridas(2023)]%
        {ebert2023generative}
\bibfield{author}{\bibinfo{person}{Christof Ebert} {and} \bibinfo{person}{Panos Louridas}.} \bibinfo{year}{2023}\natexlab{}.
\newblock \showarticletitle{{Generative AI for software practitioners}}.
\newblock \bibinfo{journal}{\emph{IEEE Software}} \bibinfo{volume}{40}, \bibinfo{number}{4} (\bibinfo{year}{2023}), \bibinfo{pages}{30--38}.
\newblock


\bibitem[Ferrara(2023)]%
        {ferrara2023fairness}
\bibfield{author}{\bibinfo{person}{Emilio Ferrara}.} \bibinfo{year}{2023}\natexlab{}.
\newblock \showarticletitle{{Fairness and Bias in Artificial Intelligence: A Brief Survey of Sources, Impacts, and Mitigation Strategies}}.
\newblock \bibinfo{journal}{\emph{Sci}} \bibinfo{volume}{6}, \bibinfo{number}{1} (\bibinfo{year}{2023}), \bibinfo{pages}{3}.
\newblock


\bibitem[figma({[n.\,d.]})]%
        {figma}
figma \bibinfo{year}{[n.\,d.]}\natexlab{}.
\newblock \bibinfo{title}{Figma}.
\newblock \bibinfo{howpublished}{\url{https://www.figma.com/}}.
\newblock


\bibitem[Filippova and Cho(2016)]%
        {filippova2016effects}
\bibfield{author}{\bibinfo{person}{Anna Filippova} {and} \bibinfo{person}{Hichang Cho}.} \bibinfo{year}{2016}\natexlab{}.
\newblock \showarticletitle{The effects and antecedents of conflict in free and open source software development}. In \bibinfo{booktitle}{\emph{Proceedings of the 19th ACM Conference on Computer-Supported Cooperative Work \& Social Computing}}. \bibinfo{pages}{705--716}.
\newblock


\bibitem[Fritsch et~al\mbox{.}(2022)]%
        {fritsch2022overview}
\bibfield{author}{\bibinfo{person}{Lothar Fritsch}, \bibinfo{person}{Aws Jaber}, {and} \bibinfo{person}{Anis Yazidi}.} \bibinfo{year}{2022}\natexlab{}.
\newblock \showarticletitle{An overview of artificial intelligence used in malware}. In \bibinfo{booktitle}{\emph{Symposium of the Norwegian AI Society}}. Springer, \bibinfo{pages}{41--51}.
\newblock


\bibitem[{Gemini Team et al.}(2023)]%
        {geminiteam2023gemini}
\bibfield{author}{\bibinfo{person}{{Gemini Team et al.}}} \bibinfo{year}{2023}\natexlab{}.
\newblock \bibinfo{title}{Gemini: A Family of Highly Capable Multimodal Models}.
\newblock
\newblock
\showeprint[arxiv]{2312.11805}~[cs.CL]


\bibitem[Generative Artificial Intelligence and Copyright Law({[n.\,d.]})]%
        {congressCopyright}
Generative Artificial Intelligence and Copyright Law \bibinfo{year}{[n.\,d.]}\natexlab{}.
\newblock \bibinfo{title}{{Generative Artificial Intelligence and Copyright Law}}.
\newblock \bibinfo{howpublished}{\url{https://crsreports.congress.gov/product/pdf/LSB/LSB10922}}.
\newblock


\bibitem[Gervais et~al\mbox{.}(2024)]%
        {gervais2024heart}
\bibfield{author}{\bibinfo{person}{Daniel~J. Gervais}, \bibinfo{person}{Noam Shemtov}, \bibinfo{person}{Haralambos Marmanis}, {and} \bibinfo{person}{Catherine Zaller~Rowland}.} \bibinfo{year}{2024}\natexlab{}.
\newblock \showarticletitle{The Heart of the Matter: Copyright, AI Training, and LLMs}.
\newblock \bibinfo{journal}{\emph{Available at SSRN, \url{https://papers.ssrn.com/sol3/papers.cfm?abstract_id=4963711}}} (\bibinfo{year}{2024}).
\newblock


\bibitem[{GitHub}({[n.\,d.]})]%
        {copilot}
\bibfield{author}{\bibinfo{person}{{GitHub}}.} \bibinfo{year}{[n.\,d.]}\natexlab{}.
\newblock \bibinfo{title}{{GitHub Copilot}}.
\newblock
\newblock
\newblock
\shownote{Retrieved March 22, 2024, from ~\url{https://copilot.github.com}}.


\bibitem[GitHub REST API documentation({[n.\,d.]})]%
        {githubAPI}
GitHub REST API documentation \bibinfo{year}{[n.\,d.]}\natexlab{}.
\newblock \bibinfo{title}{{GitHub REST API documentation}}.
\newblock \bibinfo{howpublished}{\url{https://docs.github.com/en/rest}}.
\newblock


\bibitem[Goldstein et~al\mbox{.}(2024)]%
        {goldstein2024persuasive}
\bibfield{author}{\bibinfo{person}{Josh~A Goldstein}, \bibinfo{person}{Jason Chao}, \bibinfo{person}{Shelby Grossman}, \bibinfo{person}{Alex Stamos}, {and} \bibinfo{person}{Michael Tomz}.} \bibinfo{year}{2024}\natexlab{}.
\newblock \showarticletitle{How persuasive is AI-generated propaganda?}
\newblock \bibinfo{journal}{\emph{PNAS Nexus}} \bibinfo{volume}{3}, \bibinfo{number}{2} (\bibinfo{year}{2024}), \bibinfo{pages}{pgae034}.
\newblock


\bibitem[gpt-pilot({[n.\,d.]})]%
        {gptpilot}
gpt-pilot \bibinfo{year}{[n.\,d.]}\natexlab{}.
\newblock \bibinfo{title}{gpt-pilot}.
\newblock \bibinfo{howpublished}{\url{https://github.com/Pythagora-io/gpt-pilot}}.
\newblock


\bibitem[gptengineer({[n.\,d.]})]%
        {gptengineer}
gptengineer \bibinfo{year}{[n.\,d.]}\natexlab{}.
\newblock \bibinfo{title}{gpt-engineer}.
\newblock \bibinfo{howpublished}{\url{https://github.com/gpt-engineer-org/gpt-engineer}}.
\newblock


\bibitem[grammarly({[n.\,d.]})]%
        {grammarly}
grammarly \bibinfo{year}{[n.\,d.]}\natexlab{}.
\newblock \bibinfo{title}{Transforming How the World Communicates Through AI}.
\newblock \bibinfo{howpublished}{\url{https://www.grammarly.com/ai}}.
\newblock


\bibitem[Groves et~al\mbox{.}(2009)]%
        {survey}
\bibfield{author}{\bibinfo{person}{Robert~M. Groves}, \bibinfo{person}{Floyd~J. Fowler~Jr.}, \bibinfo{person}{Mick~P. Couper}, \bibinfo{person}{James~M. Lepkowski}, \bibinfo{person}{Eleanor Singer}, {and} \bibinfo{person}{Roger Tourangeau}.} \bibinfo{year}{2009}\natexlab{}.
\newblock \bibinfo{booktitle}{\emph{{Survey Methodology, 2nd edition}}}.
\newblock \bibinfo{publisher}{Wiley}.
\newblock


\bibitem[Guadamuz(2024)]%
        {guadamuz2024scanner}
\bibfield{author}{\bibinfo{person}{Andres Guadamuz}.} \bibinfo{year}{2024}\natexlab{}.
\newblock \showarticletitle{{A Scanner Darkly: Copyright Liability and Exceptions in Artificial Intelligence Inputs and Outputs}}.
\newblock \bibinfo{journal}{\emph{GRUR International}} \bibinfo{volume}{73}, \bibinfo{number}{2} (\bibinfo{year}{2024}), \bibinfo{pages}{111--127}.
\newblock


\bibitem[Gupta et~al\mbox{.}(2023)]%
        {gupta2023chatgpt}
\bibfield{author}{\bibinfo{person}{Maanak Gupta}, \bibinfo{person}{CharanKumar Akiri}, \bibinfo{person}{Kshitiz Aryal}, \bibinfo{person}{Eli Parker}, {and} \bibinfo{person}{Lopamudra Praharaj}.} \bibinfo{year}{2023}\natexlab{}.
\newblock \showarticletitle{{From ChatGPT to ThreatGPT: Impact of Generative AI in Cybersecurity and Privacy}}.
\newblock \bibinfo{journal}{\emph{IEEE Access}} (\bibinfo{year}{2023}).
\newblock


\bibitem[He et~al\mbox{.}(2023)]%
        {he2023automating}
\bibfield{author}{\bibinfo{person}{Runzhi He}, \bibinfo{person}{Hao He}, \bibinfo{person}{Yuxia Zhang}, {and} \bibinfo{person}{Minghui Zhou}.} \bibinfo{year}{2023}\natexlab{}.
\newblock \showarticletitle{Automating dependency updates in practice: An exploratory study on github dependabot}.
\newblock \bibinfo{journal}{\emph{IEEE Transactions on Software Engineering}} \bibinfo{volume}{49}, \bibinfo{number}{8} (\bibinfo{year}{2023}), \bibinfo{pages}{4004--4022}.
\newblock


\bibitem[Henderson et~al\mbox{.}(2024)]%
        {henderson2024rethinking}
\bibfield{author}{\bibinfo{person}{Peter Henderson}, \bibinfo{person}{Jieru Hu}, \bibinfo{person}{Mona Diab}, {and} \bibinfo{person}{Joelle Pineau}.} \bibinfo{year}{2024}\natexlab{}.
\newblock \showarticletitle{{Rethinking Machine Learning Benchmarks in the Context of Professional Codes of Conduct}}. In \bibinfo{booktitle}{\emph{Proceedings of the Symposium on Computer Science and Law}}. \bibinfo{pages}{109--120}.
\newblock


\bibitem[Henderson et~al\mbox{.}(2023)]%
        {henderson2023foundation}
\bibfield{author}{\bibinfo{person}{Peter Henderson}, \bibinfo{person}{Xuechen Li}, \bibinfo{person}{Dan Jurafsky}, \bibinfo{person}{Tatsunori Hashimoto}, \bibinfo{person}{Mark~A Lemley}, {and} \bibinfo{person}{Percy Liang}.} \bibinfo{year}{2023}\natexlab{}.
\newblock \showarticletitle{{Foundation Models and Fair Use}}.
\newblock \bibinfo{journal}{\emph{arXiv preprint arXiv:2303.15715}} (\bibinfo{year}{2023}).
\newblock


\bibitem[Hong et~al\mbox{.}(2023)]%
        {hong2023metagpt}
\bibfield{author}{\bibinfo{person}{Sirui Hong}, \bibinfo{person}{Xiawu Zheng}, \bibinfo{person}{Jonathan Chen}, \bibinfo{person}{Yuheng Cheng}, \bibinfo{person}{Jinlin Wang}, \bibinfo{person}{Ceyao Zhang}, \bibinfo{person}{Zili Wang}, \bibinfo{person}{Steven Ka~Shing Yau}, \bibinfo{person}{Zijuan Lin}, \bibinfo{person}{Liyang Zhou}, {et~al\mbox{.}}} \bibinfo{year}{2023}\natexlab{}.
\newblock \showarticletitle{Metagpt: Meta programming for multi-agent collaborative framework}.
\newblock \bibinfo{journal}{\emph{arXiv preprint arXiv:2308.00352}} (\bibinfo{year}{2023}).
\newblock


\bibitem[Hood(2023)]%
        {llamafile}
\bibfield{author}{\bibinfo{person}{Stephen Hood}.} \bibinfo{year}{2023}\natexlab{}.
\newblock \bibinfo{title}{llamafile: bringing LLMs to the people, and to your own computer}.
\newblock \bibinfo{howpublished}{\url{https://future.mozilla.org/builders/news_insights/introducing-llamafile/}}.
\newblock
\newblock
\shownote{Accessed: 2024-09-11}.


\bibitem[Hu et~al\mbox{.}(2022)]%
        {hu2022membership}
\bibfield{author}{\bibinfo{person}{Hongsheng Hu}, \bibinfo{person}{Zoran Salcic}, \bibinfo{person}{Lichao Sun}, \bibinfo{person}{Gillian Dobbie}, \bibinfo{person}{Philip~S Yu}, {and} \bibinfo{person}{Xuyun Zhang}.} \bibinfo{year}{2022}\natexlab{}.
\newblock \showarticletitle{{Membership Inference Attacks on Machine Learning: A Survey}}.
\newblock \bibinfo{journal}{\emph{ACM Computing Surveys (CSUR)}} \bibinfo{volume}{54}, \bibinfo{number}{11s} (\bibinfo{year}{2022}), \bibinfo{pages}{1--37}.
\newblock


\bibitem[Huang et~al\mbox{.}(2021)]%
        {huang2021leaving}
\bibfield{author}{\bibinfo{person}{Yu Huang}, \bibinfo{person}{Denae Ford}, {and} \bibinfo{person}{Thomas Zimmermann}.} \bibinfo{year}{2021}\natexlab{}.
\newblock \showarticletitle{Leaving my fingerprints: Motivations and challenges of contributing to OSS for social good}. In \bibinfo{booktitle}{\emph{2021 IEEE/ACM 43rd International Conference on Software Engineering (ICSE)}}. IEEE, \bibinfo{pages}{1020--1032}.
\newblock


\bibitem[HuggingChat({[n.\,d.]})]%
        {huggingChat}
HuggingChat \bibinfo{year}{[n.\,d.]}\natexlab{}.
\newblock \bibinfo{title}{HuggingChat}.
\newblock \bibinfo{howpublished}{\url{https://huggingface.co/chat/}}.
\newblock
\newblock
\shownote{Accessed: 2024-08-11}.


\bibitem[Ippolito et~al\mbox{.}(2023)]%
        {ippolito2023preventing}
\bibfield{author}{\bibinfo{person}{Daphne Ippolito}, \bibinfo{person}{Florian Tram{\`e}r}, \bibinfo{person}{Milad Nasr}, \bibinfo{person}{Chiyuan Zhang}, \bibinfo{person}{Matthew Jagielski}, \bibinfo{person}{Katherine Lee}, \bibinfo{person}{Christopher~A Choquette-Choo}, {and} \bibinfo{person}{Nicholas Carlini}.} \bibinfo{year}{2023}\natexlab{}.
\newblock \showarticletitle{{Preventing Generation of Verbatim Memorization in Language Models Gives a False Sense of Privacy}}. In \bibinfo{booktitle}{\emph{Proceedings of the 16th International Natural Language Generation Conference}}. Association for Computational Linguistics, \bibinfo{pages}{28--53}.
\newblock


\bibitem[Jiang et~al\mbox{.}(2017)]%
        {jiang2017understanding}
\bibfield{author}{\bibinfo{person}{Jing Jiang}, \bibinfo{person}{David Lo}, \bibinfo{person}{Xinyu Ma}, \bibinfo{person}{Fuli Feng}, {and} \bibinfo{person}{Li Zhang}.} \bibinfo{year}{2017}\natexlab{}.
\newblock \showarticletitle{Understanding inactive yet available assignees in GitHub}.
\newblock \bibinfo{journal}{\emph{Information and Software Technology}}  \bibinfo{volume}{91} (\bibinfo{year}{2017}), \bibinfo{pages}{44--55}.
\newblock


\bibitem[Joblin et~al\mbox{.}(2017)]%
        {joblin2017classifying}
\bibfield{author}{\bibinfo{person}{Mitchell Joblin}, \bibinfo{person}{Sven Apel}, \bibinfo{person}{Claus Hunsen}, {and} \bibinfo{person}{Wolfgang Mauerer}.} \bibinfo{year}{2017}\natexlab{}.
\newblock \showarticletitle{Classifying developers into core and peripheral: An empirical study on count and network metrics}. In \bibinfo{booktitle}{\emph{2017 IEEE/ACM 39th International Conference on Software Engineering (ICSE)}}. IEEE, \bibinfo{pages}{164--174}.
\newblock


\bibitem[Karathanasis(2023)]%
        {FrenchCase2023}
\bibfield{author}{\bibinfo{person}{Theodoros Karathanasis}.} \bibinfo{year}{2023}\natexlab{}.
\newblock \bibinfo{title}{{EU Copyright Directive: A `Nightmare' for Generative AI Researchers and Developers?}}
\newblock
\newblock
\newblock
\shownote{\url{https://ai-regulation.com/eu-copyright-directive-a-nightmare-for-gai/}}.


\bibitem[Kitchenham and Pfleeger(2002a)]%
        {DBLP:journals/sigsoft/KitchenhamP02}
\bibfield{author}{\bibinfo{person}{Barbara~A. Kitchenham} {and} \bibinfo{person}{Shari~Lawrence Pfleeger}.} \bibinfo{year}{2002}\natexlab{a}.
\newblock \showarticletitle{{Principles of Survey Research Part 2: Designing a Survey}}.
\newblock \bibinfo{journal}{\emph{{ACM} {SIGSOFT} Software Engineering Notes}} \bibinfo{volume}{27}, \bibinfo{number}{1} (\bibinfo{year}{2002}), \bibinfo{pages}{18--20}.
\newblock


\bibitem[Kitchenham and Pfleeger(2002b)]%
        {DBLP:journals/sigsoft/KitchenhamP02a}
\bibfield{author}{\bibinfo{person}{Barbara~A. Kitchenham} {and} \bibinfo{person}{Shari~Lawrence Pfleeger}.} \bibinfo{year}{2002}\natexlab{b}.
\newblock \showarticletitle{{Principles of Survey Research: Part 3: Constructing a Survey Instrument}}.
\newblock \bibinfo{journal}{\emph{{ACM} {SIGSOFT} Software Engineering Notes}} \bibinfo{volume}{27}, \bibinfo{number}{2} (\bibinfo{year}{2002}), \bibinfo{pages}{20--24}.
\newblock


\bibitem[Kitchenham and Pfleeger(2002c)]%
        {DBLP:journals/sigsoft/KitchenhamP02b}
\bibfield{author}{\bibinfo{person}{Barbara~A. Kitchenham} {and} \bibinfo{person}{Shari~Lawrence Pfleeger}.} \bibinfo{year}{2002}\natexlab{c}.
\newblock \showarticletitle{{Principles of Survey Research Part 4: Questionnaire Evaluation}}.
\newblock \bibinfo{journal}{\emph{{ACM} {SIGSOFT} Software Engineering Notes}} \bibinfo{volume}{27}, \bibinfo{number}{3} (\bibinfo{year}{2002}), \bibinfo{pages}{20--23}.
\newblock


\bibitem[Kitchenham and Pfleeger(2002d)]%
        {DBLP:journals/sigsoft/KitchenhamP02c}
\bibfield{author}{\bibinfo{person}{Barbara~A. Kitchenham} {and} \bibinfo{person}{Shari~Lawrence Pfleeger}.} \bibinfo{year}{2002}\natexlab{d}.
\newblock \showarticletitle{{Principles of Survey Research: Part 5: Populations and Samples}}.
\newblock \bibinfo{journal}{\emph{{ACM} {SIGSOFT} Software Engineering Notes}} \bibinfo{volume}{27}, \bibinfo{number}{5} (\bibinfo{year}{2002}), \bibinfo{pages}{17--20}.
\newblock


\bibitem[Kitchenham and Pfleeger(2003)]%
        {DBLP:journals/sigsoft/KitchenhamP03}
\bibfield{author}{\bibinfo{person}{Barbara~A. Kitchenham} {and} \bibinfo{person}{Shari~Lawrence Pfleeger}.} \bibinfo{year}{2003}\natexlab{}.
\newblock \showarticletitle{{Principles of Survey Research Part 6: Data Analysis}}.
\newblock \bibinfo{journal}{\emph{{ACM} {SIGSOFT} Software Engineering Notes}} \bibinfo{volume}{28}, \bibinfo{number}{2} (\bibinfo{year}{2003}), \bibinfo{pages}{24--27}.
\newblock


\bibitem[Krill({[n.\,d.]})]%
        {all_using_ai}
\bibfield{author}{\bibinfo{person}{Paul Krill}.} \bibinfo{year}{[n.\,d.]}\natexlab{}.
\newblock \bibinfo{title}{GitHub survey finds nearly all developers using AI coding tools}.
\newblock \bibinfo{howpublished}{\url{https://www.infoworld.com/article/3489925/github-survey-finds-nearly-all-developers-using-ai-coding-tools.html}}.
\newblock


\bibitem[Kugler(2024)]%
        {Kugler2024WhoOA}
\bibfield{author}{\bibinfo{person}{Logan Kugler}.} \bibinfo{year}{2024}\natexlab{}.
\newblock \showarticletitle{Who Owns AI’s Output?}
\newblock \bibinfo{journal}{\emph{Commun. ACM}} (\bibinfo{year}{2024}).
\newblock
\urldef\tempurl%
\url{https://api.semanticscholar.org/CorpusID:273132452}
\showURL{%
\tempurl}


\bibitem[Lanz({[n.\,d.]})]%
        {japanAI}
\bibfield{author}{\bibinfo{person}{Jose~Antonio Lanz}.} \bibinfo{year}{[n.\,d.]}\natexlab{}.
\newblock \bibinfo{title}{{AI Art Wars: Japan Says AI Model Training Doesn’t Violate Copyright}}.
\newblock \bibinfo{howpublished}{https://finance.yahoo.com/news/ai-art-wars-japan-says-185350499.html}.
\newblock


\bibitem[Lee et~al\mbox{.}(2023a)]%
        {lee2023talkin}
\bibfield{author}{\bibinfo{person}{Katherine Lee}, \bibinfo{person}{A.~Feder Cooper}, {and} \bibinfo{person}{James Grimmelmann}.} \bibinfo{year}{2023}\natexlab{a}.
\newblock \showarticletitle{{Talkin' 'Bout {AI} Generation: Copyright and the Generative-{AI} Supply Chain}}.
\newblock \bibinfo{journal}{\emph{arXiv preprint arXiv:2309.08133}} (\bibinfo{year}{2023}).
\newblock


\bibitem[Lee et~al\mbox{.}(2023b)]%
        {devilInDetails}
\bibfield{author}{\bibinfo{person}{Katherine Lee}, \bibinfo{person}{A.~Feder Cooper}, \bibinfo{person}{James Grimmelmann}, {and} \bibinfo{person}{Daphne Ippolito}.} \bibinfo{year}{2023}\natexlab{b}.
\newblock \bibinfo{title}{{The Devil Is in the Training Data}}.
\newblock \bibinfo{howpublished}{\url{https://genlaw.org/explainers/training-data.html}}.
\newblock


\bibitem[Lemley and Casey(2020)]%
        {lemley2020fair}
\bibfield{author}{\bibinfo{person}{Mark~A. Lemley} {and} \bibinfo{person}{Bryan Casey}.} \bibinfo{year}{2020}\natexlab{}.
\newblock \showarticletitle{Fair Learning}.
\newblock \bibinfo{journal}{\emph{Tex. L. Rev.}}  \bibinfo{volume}{99} (\bibinfo{year}{2020}), \bibinfo{pages}{743--785}.
\newblock


\bibitem[Li and Ahmed(2023)]%
        {li2023commit}
\bibfield{author}{\bibinfo{person}{Jiawei Li} {and} \bibinfo{person}{Iftekhar Ahmed}.} \bibinfo{year}{2023}\natexlab{}.
\newblock \showarticletitle{Commit message matters: Investigating impact and evolution of commit message quality}. In \bibinfo{booktitle}{\emph{2023 IEEE/ACM 45th International Conference on Software Engineering (ICSE)}}. IEEE, \bibinfo{pages}{806--817}.
\newblock


\bibitem[Liang et~al\mbox{.}(2024)]%
        {liang2024large}
\bibfield{author}{\bibinfo{person}{Jenny~T Liang}, \bibinfo{person}{Chenyang Yang}, {and} \bibinfo{person}{Brad~A Myers}.} \bibinfo{year}{2024}\natexlab{}.
\newblock \showarticletitle{A large-scale survey on the usability of {AI} programming assistants: Successes and challenges}. In \bibinfo{booktitle}{\emph{Proceedings of the 46th IEEE/ACM International Conference on Software Engineering}}. \bibinfo{pages}{1--13}.
\newblock


\bibitem[Liang et~al\mbox{.}(2022)]%
        {liang2022understanding}
\bibfield{author}{\bibinfo{person}{Jenny~T Liang}, \bibinfo{person}{Thomas Zimmermann}, {and} \bibinfo{person}{Denae Ford}.} \bibinfo{year}{2022}\natexlab{}.
\newblock \showarticletitle{Understanding skills for OSS communities on GitHub}. In \bibinfo{booktitle}{\emph{Proceedings of the 30th ACM Joint European Software Engineering Conference and Symposium on the Foundations of Software Engineering}}. \bibinfo{pages}{170--182}.
\newblock


\bibitem[Liu et~al\mbox{.}(2024)]%
        {liu2024hitchhiker}
\bibfield{author}{\bibinfo{person}{Yi Liu}, \bibinfo{person}{Gelei Deng}, \bibinfo{person}{Zhengzi Xu}, \bibinfo{person}{Yuekang Li}, \bibinfo{person}{Yaowen Zheng}, \bibinfo{person}{Ying Zhang}, \bibinfo{person}{Lida Zhao}, \bibinfo{person}{Tianwei Zhang}, {and} \bibinfo{person}{Kailong Wang}.} \bibinfo{year}{2024}\natexlab{}.
\newblock \showarticletitle{A Hitchhiker’s Guide to Jailbreaking ChatGPT via Prompt Engineering}. In \bibinfo{booktitle}{\emph{Proceedings of the 4th International Workshop on Software Engineering and AI for Data Quality in Cyber-Physical Systems/Internet of Things}}. \bibinfo{pages}{12--21}.
\newblock


\bibitem[Llama.cpp({[n.\,d.]})]%
        {llamacpp}
Llama.cpp \bibinfo{year}{[n.\,d.]}\natexlab{}.
\newblock \bibinfo{title}{Llama.cpp}.
\newblock \bibinfo{howpublished}{\url{https://github.com/ggerganov/llama.cpp}}.
\newblock
\newblock
\shownote{Accessed: 2024-08-11}.


\bibitem[Lucchi(2023)]%
        {lucchi2023chatgpt}
\bibfield{author}{\bibinfo{person}{Nicola Lucchi}.} \bibinfo{year}{2023}\natexlab{}.
\newblock \showarticletitle{{ChatGPT: A Case Study on Copyright Challenges for Generative Artificial Intelligence Systems}}.
\newblock \bibinfo{journal}{\emph{European Journal of Risk Regulation}} (\bibinfo{year}{2023}), \bibinfo{pages}{1--23}.
\newblock


\bibitem[Ma et~al\mbox{.}(2024)]%
        {ma2024crafting}
\bibfield{author}{\bibinfo{person}{Lijia Ma}, \bibinfo{person}{Xingchen Xu}, {and} \bibinfo{person}{Yong Tan}.} \bibinfo{year}{2024}\natexlab{}.
\newblock \showarticletitle{{Crafting Knowledge: Exploring the Creative Mechanisms of Chat-Based Search Engines}}.
\newblock \bibinfo{journal}{\emph{arXiv preprint arXiv:2402.19421}} (\bibinfo{year}{2024}).
\newblock


\bibitem[Majdinasab et~al\mbox{.}(2024a)]%
        {majdinasab2024assessing}
\bibfield{author}{\bibinfo{person}{Vahid Majdinasab}, \bibinfo{person}{Michael~Joshua Bishop}, \bibinfo{person}{Shawn Rasheed}, \bibinfo{person}{Arghavan Moradidakhel}, \bibinfo{person}{Amjed Tahir}, {and} \bibinfo{person}{Foutse Khomh}.} \bibinfo{year}{2024}\natexlab{a}.
\newblock \showarticletitle{Assessing the Security of GitHub Copilot's Generated Code-A Targeted Replication Study}. In \bibinfo{booktitle}{\emph{2024 IEEE International Conference on Software Analysis, Evolution and Reengineering (SANER)}}. IEEE, \bibinfo{pages}{435--444}.
\newblock


\bibitem[Majdinasab et~al\mbox{.}(2024b)]%
        {majdinasab2024trained}
\bibfield{author}{\bibinfo{person}{Vahid Majdinasab}, \bibinfo{person}{Amin Nikanjam}, {and} \bibinfo{person}{Foutse Khomh}.} \bibinfo{year}{2024}\natexlab{b}.
\newblock \showarticletitle{{Trained Without My Consent: Detecting Code Inclusion in Language Models Trained on Code}}.
\newblock \bibinfo{journal}{\emph{arXiv preprint arXiv:2402.09299}} (\bibinfo{year}{2024}).
\newblock


\bibitem[Malley(2023)]%
        {chatbotProblems}
\bibfield{author}{\bibinfo{person}{Tom Malley}.} \bibinfo{year}{2023}\natexlab{}.
\newblock \bibinfo{title}{{AI Have a Deal: Driver uses ChatGPT} hack to get dealer to agree to sell new car for \$1 in ‘legally binding deal’ in blow for {AI} rollout}.
\newblock \bibinfo{howpublished}{\url{https://www.the-sun.com/motors/9888857/driver-uses-ai-loophole-buy-new-car-1}}.
\newblock


\bibitem[May({[n.\,d.]})]%
        {stackoverflow_survey}
\bibfield{author}{\bibinfo{person}{Eira May}.} \bibinfo{year}{[n.\,d.]}\natexlab{}.
\newblock \bibinfo{title}{{Where developers feel AI coding tools are working—and where they’re missing the mark}}.
\newblock \bibinfo{howpublished}{\url{https://stackoverflow.blog/2024/09/23/where-developers-feel-ai-coding-tools-are-working-and-where-they-re-missing-the-mark/}}.
\newblock


\bibitem[McCallum(2023)]%
        {ChatGPTItalyBan}
\bibfield{author}{\bibinfo{person}{Shiona McCallum}.} \bibinfo{year}{2023}\natexlab{}.
\newblock \bibinfo{title}{{ChatGPT banned in Italy over privacy concerns}}.
\newblock
\newblock
\newblock
\shownote{\url{https://www.bbc.com/news/technology-65139406}}.


\bibitem[Meta(2024)]%
        {llama3_release}
\bibfield{author}{\bibinfo{person}{Meta}.} \bibinfo{year}{2024}\natexlab{}.
\newblock \bibinfo{title}{Introducing Meta Llama 3: The most capable openly available LLM to date}.
\newblock \bibinfo{howpublished}{\url{https://ai.meta.com/blog/meta-llama-3/}}.
\newblock


\bibitem[Mitchell et~al\mbox{.}(2019)]%
        {mitchell2019model}
\bibfield{author}{\bibinfo{person}{Margaret Mitchell}, \bibinfo{person}{Simone Wu}, \bibinfo{person}{Andrew Zaldivar}, \bibinfo{person}{Parker Barnes}, \bibinfo{person}{Lucy Vasserman}, \bibinfo{person}{Ben Hutchinson}, \bibinfo{person}{Elena Spitzer}, \bibinfo{person}{Inioluwa~Deborah Raji}, {and} \bibinfo{person}{Timnit Gebru}.} \bibinfo{year}{2019}\natexlab{}.
\newblock \showarticletitle{{Model Cards for Model Reporting}}. In \bibinfo{booktitle}{\emph{Proceedings of the conference on fairness, accountability, and transparency}}. \bibinfo{pages}{220--229}.
\newblock


\bibitem[Moraes et~al\mbox{.}(2021)]%
        {moraes2021one}
\bibfield{author}{\bibinfo{person}{Joao~Pedro Moraes}, \bibinfo{person}{Ivanilton Polato}, \bibinfo{person}{Igor Wiese}, \bibinfo{person}{Filipe Saraiva}, {and} \bibinfo{person}{Gustavo Pinto}.} \bibinfo{year}{2021}\natexlab{}.
\newblock \showarticletitle{From one to hundreds: multi-licensing in the JavaScript ecosystem}.
\newblock \bibinfo{journal}{\emph{Empirical Software Engineering}}  \bibinfo{volume}{26} (\bibinfo{year}{2021}), \bibinfo{pages}{1--29}.
\newblock


\bibitem[Neel and Chang(2023)]%
        {neel2023privacy}
\bibfield{author}{\bibinfo{person}{Seth Neel} {and} \bibinfo{person}{Peter Chang}.} \bibinfo{year}{2023}\natexlab{}.
\newblock \showarticletitle{{Privacy Issues in Large Language Models: A Survey}}.
\newblock \bibinfo{journal}{\emph{arXiv preprint arXiv:2312.06717}} (\bibinfo{year}{2023}).
\newblock


\bibitem[Newsroom(2023)]%
        {wormgpt}
\bibfield{author}{\bibinfo{person}{Newsroom}.} \bibinfo{year}{2023}\natexlab{}.
\newblock \bibinfo{title}{{WormGPT}: New {AI} Tool Allows Cybercriminals to Launch Sophisticated Cyber Attacks}.
\newblock \bibinfo{howpublished}{\url{https://thehackernews.com/2023/07/wormgpt-new-ai-tool-allows.html}}.
\newblock


\bibitem[Office(2023a)]%
        {copyright_office_NOI}
\bibfield{author}{\bibinfo{person}{United States~Copyright Office}.} \bibinfo{year}{2023}\natexlab{a}.
\newblock \bibinfo{title}{Artificial Intelligence and Copyright}.
\newblock \bibinfo{howpublished}{\url{https://www.federalregister.gov/documents/2023/08/30/2023-18624/artificial-intelligence-and-copyright}}.
\newblock


\bibitem[Office(2023b)]%
        {usCopyRegis23}
\bibfield{author}{\bibinfo{person}{United States~Copyright Office}.} \bibinfo{year}{2023}\natexlab{b}.
\newblock \bibinfo{title}{Copyright Registration Guidance: Works Containing Material Generated by Artificial Intelligence}.
\newblock
\newblock
\newblock
\shownote{16190 Federal Register, Vol. 88, No. 51}.


\bibitem[Office(2024a)]%
        {UCOAIstudy}
\bibfield{author}{\bibinfo{person}{United States~Copyright Office}.} \bibinfo{year}{2024}\natexlab{a}.
\newblock \bibinfo{title}{Artificial Intelligence Study}.
\newblock \bibinfo{howpublished}{\url{https://www.copyright.gov/policy/artificial-intelligence/}}.
\newblock


\bibitem[Office(2024b)]%
        {copyright_office_1}
\bibfield{author}{\bibinfo{person}{United States~Copyright Office}.} \bibinfo{year}{2024}\natexlab{b}.
\newblock \bibinfo{title}{Copyright and Artificial Intelligence Part 1: Digital Replicas}.
\newblock \bibinfo{howpublished}{\url{https://www.copyright.gov/ai/Copyright-and-Artificial-Intelligence-Part-1-Digital-Replicas-Report.pdf}}.
\newblock


\bibitem[Office(2025)]%
        {copyright_office_2}
\bibfield{author}{\bibinfo{person}{United States~Copyright Office}.} \bibinfo{year}{2025}\natexlab{}.
\newblock \bibinfo{title}{Copyright and Artificial Intelligence Part 2: Copyrightability}.
\newblock \bibinfo{howpublished}{\url{https://www.copyright.gov/ai/Copyright-and-Artificial-Intelligence-Part-2-Copyrightability-Report.pdf}}.
\newblock


\bibitem[Ollama({[n.\,d.]})]%
        {ollama}
Ollama \bibinfo{year}{[n.\,d.]}\natexlab{}.
\newblock \bibinfo{title}{Ollama}.
\newblock \bibinfo{howpublished}{\url{https://ollama.com/}}.
\newblock
\newblock
\shownote{Accessed: 2024-08-11}.


\bibitem[Oobabooga({[n.\,d.]})]%
        {oobabooga}
Oobabooga \bibinfo{year}{[n.\,d.]}\natexlab{}.
\newblock \bibinfo{title}{Oobabooga Text Generation WebUI}.
\newblock \bibinfo{howpublished}{\url{https://github.com/oobabooga/text-generation-webui}}.
\newblock
\newblock
\shownote{Accessed: 2024-08-11}.


\bibitem[{OpenAI}({[n.\,d.]})]%
        {chatgpt}
\bibfield{author}{\bibinfo{person}{{OpenAI}}.} \bibinfo{year}{[n.\,d.]}\natexlab{}.
\newblock \bibinfo{title}{{ChatGPT}~\url{https://openai.com/blog/chatgpt}}.
\newblock
\newblock
\newblock
\shownote{Last accessed: March 2024}.


\bibitem[OpenAI(2024)]%
        {openai_guardrails}
\bibfield{author}{\bibinfo{person}{OpenAI}.} \bibinfo{year}{2024}\natexlab{}.
\newblock \bibinfo{title}{OpenAI safety update}.
\newblock \bibinfo{howpublished}{\url{https://openai.com/index/openai-safety-update/}}.
\newblock


\bibitem[Oppenheim(2000)]%
        {oppenheim2000questionnaire}
\bibfield{author}{\bibinfo{person}{Abraham~Naftali Oppenheim}.} \bibinfo{year}{2000}\natexlab{}.
\newblock \bibinfo{booktitle}{\emph{Questionnaire design, interviewing and attitude measurement}}.
\newblock \bibinfo{publisher}{Bloomsbury Publishing}.
\newblock


\bibitem[Pa~Pa et~al\mbox{.}(2023)]%
        {pa2023attacker}
\bibfield{author}{\bibinfo{person}{Yin~Minn Pa~Pa}, \bibinfo{person}{Shunsuke Tanizaki}, \bibinfo{person}{Tetsui Kou}, \bibinfo{person}{Michel Van~Eeten}, \bibinfo{person}{Katsunari Yoshioka}, {and} \bibinfo{person}{Tsutomu Matsumoto}.} \bibinfo{year}{2023}\natexlab{}.
\newblock \showarticletitle{An attacker’s dream? exploring the capabilities of chatgpt for developing malware}. In \bibinfo{booktitle}{\emph{Proceedings of the 16th Cyber Security Experimentation and Test Workshop}}. \bibinfo{pages}{10--18}.
\newblock


\bibitem[Perplexity({[n.\,d.]})]%
        {perplexity}
Perplexity \bibinfo{year}{[n.\,d.]}\natexlab{}.
\newblock \bibinfo{title}{What is Perplexity?}
\newblock \bibinfo{howpublished}{\url{https://www.perplexity.ai/hub/faq/what-is-perplexity}}.
\newblock
\newblock
\shownote{Accessed: 2024-08-11}.


\bibitem[Pfleeger and Kitchenham(2001)]%
        {DBLP:journals/sigsoft/PfleegerK01}
\bibfield{author}{\bibinfo{person}{Shari~Lawrence Pfleeger} {and} \bibinfo{person}{Barbara~A. Kitchenham}.} \bibinfo{year}{2001}\natexlab{}.
\newblock \showarticletitle{{Principles of Survey Research: Part 1: Turning Lemons into Lemonade}}.
\newblock \bibinfo{journal}{\emph{{ACM} {SIGSOFT} Software Engineering Notes}} \bibinfo{volume}{26}, \bibinfo{number}{6} (\bibinfo{year}{2001}), \bibinfo{pages}{16--18}.
\newblock


\bibitem[{R. Li et al.}(2023)]%
        {li2023starcoder}
\bibfield{author}{\bibinfo{person}{{R. Li et al.}}} \bibinfo{year}{2023}\natexlab{}.
\newblock \showarticletitle{{StarCoder: may the source be with you}}.
\newblock \bibinfo{journal}{\emph{arXiv preprint arXiv:2305.06161}} (\bibinfo{year}{2023}).
\newblock


\bibitem[Radford et~al\mbox{.}(2023)]%
        {radford2023robust}
\bibfield{author}{\bibinfo{person}{Alec Radford}, \bibinfo{person}{Jong~Wook Kim}, \bibinfo{person}{Tao Xu}, \bibinfo{person}{Greg Brockman}, \bibinfo{person}{Christine McLeavey}, {and} \bibinfo{person}{Ilya Sutskever}.} \bibinfo{year}{2023}\natexlab{}.
\newblock \showarticletitle{Robust speech recognition via large-scale weak supervision}. In \bibinfo{booktitle}{\emph{International conference on machine learning}}. PMLR, \bibinfo{pages}{28492--28518}.
\newblock


\bibitem[Rajbhoj et~al\mbox{.}(2024)]%
        {rajbhoj2024accelerating}
\bibfield{author}{\bibinfo{person}{Asha Rajbhoj}, \bibinfo{person}{Akanksha Somase}, \bibinfo{person}{Piyush Kulkarni}, {and} \bibinfo{person}{Vinay Kulkarni}.} \bibinfo{year}{2024}\natexlab{}.
\newblock \showarticletitle{{Accelerating Software Development Using Generative AI: ChatGPT Case Study}}. In \bibinfo{booktitle}{\emph{Proceedings of the 17th Innovations in Software Engineering Conference}}. \bibinfo{pages}{1--11}.
\newblock


\bibitem[{Responsible AI}(2022)]%
        {openrail}
\bibfield{author}{\bibinfo{person}{{Responsible AI}}.} \bibinfo{year}{2022}\natexlab{}.
\newblock \bibinfo{title}{{Big Science Open Rail-M License}~\url{https://www.licenses.ai/blog/2022/8/26/bigscience-open-rail-m-license}}.
\newblock
\newblock


\bibitem[Sarkari(2024)]%
        {dolphin_uncensored}
\bibfield{author}{\bibinfo{person}{Aresh Sarkari}.} \bibinfo{year}{2024}\natexlab{}.
\newblock \bibinfo{title}{Exploring Uncensored LLM Model – Dolphin 2.9 on Llama-3-8b}.
\newblock \bibinfo{howpublished}{\url{https://askaresh.com/2024/05/02/exploring-uncensored-llm-model-dolphin-2-9-on-llama-3-8b/}}.
\newblock


\bibitem[Saroar and Nayebi(2023)]%
        {saroar2023developers}
\bibfield{author}{\bibinfo{person}{Sk~Golam Saroar} {and} \bibinfo{person}{Maleknaz Nayebi}.} \bibinfo{year}{2023}\natexlab{}.
\newblock \showarticletitle{Developers’ perception of GitHub Actions: A survey analysis}. In \bibinfo{booktitle}{\emph{Proceedings of the 27th International Conference on Evaluation and Assessment in Software Engineering}}. \bibinfo{pages}{121--130}.
\newblock


\bibitem[Sergeyuk et~al\mbox{.}(2024)]%
        {abs-2406-07765}
\bibfield{author}{\bibinfo{person}{Agnia Sergeyuk}, \bibinfo{person}{Yaroslav Golubev}, \bibinfo{person}{Timofey Bryksin}, {and} \bibinfo{person}{Iftekhar Ahmed}.} \bibinfo{year}{2024}\natexlab{}.
\newblock \showarticletitle{Using AI-Based Coding Assistants in Practice: State of Affairs, Perceptions, and Ways Forward}.
\newblock \bibinfo{journal}{\emph{CoRR}}  \bibinfo{volume}{abs/2406.07765} (\bibinfo{year}{2024}).
\newblock


\bibitem[Shinbun(2023)]%
        {JapanNewsAIPolicy23}
\bibfield{author}{\bibinfo{person}{The~Yomieru Shinbun}.} \bibinfo{year}{June 10, 2023}\natexlab{}.
\newblock \bibinfo{title}{{Intellectual Property Plan Signals Reversal on AI Policy}}.
\newblock \bibinfo{howpublished}{Japan News}.
\newblock


\bibitem[Spencer(2009)]%
        {spencer2009card}
\bibfield{author}{\bibinfo{person}{Donna Spencer}.} \bibinfo{year}{2009}\natexlab{}.
\newblock \bibinfo{booktitle}{\emph{Card sorting: Designing usable categories}}.
\newblock \bibinfo{publisher}{Rosenfeld Media}.
\newblock


\bibitem[Stallman({[n.\,d.]})]%
        {whygplv3}
\bibfield{author}{\bibinfo{person}{Richard Stallman}.} \bibinfo{year}{[n.\,d.]}\natexlab{}.
\newblock \bibinfo{title}{{Why Upgrade to GPLv3}}.
\newblock \bibinfo{howpublished}{\url{https://www.gnu.org/licenses/rms-why-gplv3}}.
\newblock


\bibitem[Stalnaker et~al\mbox{.}(2024)]%
        {stalnaker2024boms}
\bibfield{author}{\bibinfo{person}{Trevor Stalnaker}, \bibinfo{person}{Nathan Wintersgill}, \bibinfo{person}{Oscar Chaparro}, \bibinfo{person}{Massimiliano Di~Penta}, \bibinfo{person}{Daniel~M German}, {and} \bibinfo{person}{Denys Poshyvanyk}.} \bibinfo{year}{2024}\natexlab{}.
\newblock \showarticletitle{{BOMs Away! Inside the Minds of Stakeholders: A Comprehensive Study of Bills of Materials for Software Systems}}. In \bibinfo{booktitle}{\emph{Proceedings of the 46th IEEE/ACM International Conference on Software Engineering}}. \bibinfo{pages}{1--13}.
\newblock


\bibitem[tabby({[n.\,d.]})]%
        {tabby}
tabby \bibinfo{year}{[n.\,d.]}\natexlab{}.
\newblock \bibinfo{title}{tabby}.
\newblock \bibinfo{howpublished}{\url{https://github.com/TabbyML/tabby}}.
\newblock


\bibitem[Tabnine({[n.\,d.]})]%
        {tabnine}
Tabnine \bibinfo{year}{[n.\,d.]}\natexlab{}.
\newblock \bibinfo{title}{The AI code assistant you control}.
\newblock \bibinfo{howpublished}{\url{https://www.tabnine.com/}}.
\newblock
\newblock
\shownote{Accessed: 2024-25-10}.


\bibitem[Team(2021)]%
        {whyagpl}
\bibfield{author}{\bibinfo{person}{FOSSA~Editorial Team}.} \bibinfo{year}{2021}\natexlab{}.
\newblock \bibinfo{title}{{Open Source Software Licenses 101: The AGPL License}}.
\newblock \bibinfo{howpublished}{\url{https://fossa.com/blog/open-source-software-licenses-101-agpl-license/}}.
\newblock


\bibitem[The New York Times Company v. Microsoft Corporation({[n.\,d.]})]%
        {2023chatgpt}
The New York Times Company v. Microsoft Corporation \bibinfo{year}{[n.\,d.]}\natexlab{}.
\newblock \bibinfo{howpublished}{The New York Times Company v. Microsoft Corporation, No. 1:23-cv-11195 (S.D.N.Y., filed Dec. 27, 2023), \url{https://nytco-assets.nytimes.com/2023/12/NYT_Complaint_Dec2023.pdf}}.
\newblock


\bibitem[Wei et~al\mbox{.}(2024)]%
        {wei2024jailbroken}
\bibfield{author}{\bibinfo{person}{Alexander Wei}, \bibinfo{person}{Nika Haghtalab}, {and} \bibinfo{person}{Jacob Steinhardt}.} \bibinfo{year}{2024}\natexlab{}.
\newblock \showarticletitle{Jailbroken: How does LLM safety training fail?}
\newblock \bibinfo{journal}{\emph{Advances in Neural Information Processing Systems}}  \bibinfo{volume}{36} (\bibinfo{year}{2024}).
\newblock


\bibitem[Whisper({[n.\,d.]})]%
        {whisperRepo}
Whisper \bibinfo{year}{[n.\,d.]}\natexlab{}.
\newblock \bibinfo{title}{Whisper}.
\newblock \bibinfo{howpublished}{\url{https://github.com/openai/whisper}}.
\newblock
\newblock
\shownote{Accessed: 2024-11-12}.


\bibitem[Wintersgill et~al\mbox{.}(2024)]%
        {wintersgill2024law}
\bibfield{author}{\bibinfo{person}{Nathan Wintersgill}, \bibinfo{person}{Trevor Stalnaker}, \bibinfo{person}{Laura~A. Heymann}, \bibinfo{person}{Oscar Chaparro}, {and} \bibinfo{person}{Denys Poshyvanyk}.} \bibinfo{year}{2024}\natexlab{}.
\newblock \showarticletitle{{“The Law Doesn’t Work Like a Computer”: Exploring Software Licensing Issues Faced by Legal Practitioners}}. In \bibinfo{booktitle}{\emph{Proceedings of the ACM on Software Engineering}}, Vol.~\bibinfo{volume}{1}. \bibinfo{publisher}{ACM New York, NY, USA}, \bibinfo{pages}{882--905}.
\newblock


\bibitem[Wu(2024)]%
        {devin}
\bibfield{author}{\bibinfo{person}{Scott Wu}.} \bibinfo{year}{2024}\natexlab{}.
\newblock \bibinfo{title}{{Meet Devin: The World’s First AI Software Engineer}}.
\newblock \bibinfo{howpublished}{\url{https://www.cognition-labs.com/introducing-devin}}.
\newblock


\bibitem[XAI(2024)]%
        {xai_grok}
\bibfield{author}{\bibinfo{person}{XAI}.} \bibinfo{year}{2024}\natexlab{}.
\newblock \bibinfo{title}{Open Release of Grok-1}.
\newblock \bibinfo{howpublished}{\url{https://x.ai/blog/grok-os}}.
\newblock


\bibitem[Xia et~al\mbox{.}(2023)]%
        {xia2023empirical}
\bibfield{author}{\bibinfo{person}{Boming Xia}, \bibinfo{person}{Tingting Bi}, \bibinfo{person}{Zhenchang Xing}, \bibinfo{person}{Qinghua Lu}, {and} \bibinfo{person}{Liming Zhu}.} \bibinfo{year}{2023}\natexlab{}.
\newblock \showarticletitle{{An Empirical Study on Software Bill of Materials: Where We Stand and the Road Ahead}}. In \bibinfo{booktitle}{\emph{2023 IEEE/ACM 45th International Conference on Software Engineering (ICSE)}}. IEEE, \bibinfo{pages}{2630--2642}.
\newblock


\bibitem[Yang et~al\mbox{.}(2024)]%
        {yang2024unveiling}
\bibfield{author}{\bibinfo{person}{Zhou Yang}, \bibinfo{person}{Zhipeng Zhao}, \bibinfo{person}{Chenyu Wang}, \bibinfo{person}{Jieke Shi}, \bibinfo{person}{Dongsun Kim}, \bibinfo{person}{Donggyun Han}, {and} \bibinfo{person}{David Lo}.} \bibinfo{year}{2024}\natexlab{}.
\newblock \showarticletitle{{Unveiling Memorization in Code Models}}. In \bibinfo{booktitle}{\emph{2024 IEEE/ACM 46th International Conference on Software Engineering (ICSE)}}. IEEE Computer Society, \bibinfo{pages}{856--856}.
\newblock


\bibitem[Yao(2023)]%
        {copilotAdoption}
\bibfield{author}{\bibinfo{person}{Deborah Yao}.} \bibinfo{year}{2023}\natexlab{}.
\newblock \showarticletitle{{One Year On, GitHub Copilot Adoption Soars}}.
\newblock \bibinfo{howpublished}{\url{https://aibusiness.com/companies/one-year-on-github-copilot-adoption-soars}}.
\newblock \bibinfo{journal}{\emph{AI Business}} (\bibinfo{date}{27 6} \bibinfo{year}{2023}).
\newblock
\newblock
\shownote{Accessed: March 22, 2024. \url{https://aibusiness.com/companies/one-year-on-github-copilot-adoption-soars}}.


\bibitem[Yao et~al\mbox{.}(2023)]%
        {yao2023}
\bibfield{author}{\bibinfo{person}{Yuanshun Yao}, \bibinfo{person}{Xiaojun Xu}, {and} \bibinfo{person}{Yang Liu}.} \bibinfo{year}{2023}\natexlab{}.
\newblock \showarticletitle{{Large Language Model Unlearning}}.
\newblock \bibinfo{journal}{\emph{CoRR}}  \bibinfo{volume}{abs/2310.10683} (\bibinfo{year}{2023}).
\newblock


\bibitem[Yew(2024)]%
        {yew2024break}
\bibfield{author}{\bibinfo{person}{Rui-Jie Yew}.} \bibinfo{year}{2024}\natexlab{}.
\newblock \showarticletitle{{Break It 'Til You Make It: An Exploration of the Ramifications of Copyright Liability Under a Pre-training Paradigm of {AI} Development}}. In \bibinfo{booktitle}{\emph{Proceedings of the Symposium on Computer Science and Law}}. \bibinfo{pages}{64--72}.
\newblock


\bibitem[{Z. Yang et al.}(2023)]%
        {yang2023gotcha}
\bibfield{author}{\bibinfo{person}{{Z. Yang et al.}}} \bibinfo{year}{2023}\natexlab{}.
\newblock \showarticletitle{{Gotcha! This Model Uses My Code! Evaluating Membership Leakage Risks in Code Models}}.
\newblock \bibinfo{journal}{\emph{arXiv preprint arXiv:2310.01166}} (\bibinfo{year}{2023}).
\newblock


\bibitem[Zhang and Li(2023)]%
        {zhang2023code}
\bibfield{author}{\bibinfo{person}{Sheng Zhang} {and} \bibinfo{person}{Hui Li}.} \bibinfo{year}{2023}\natexlab{}.
\newblock \showarticletitle{{Code Membership Inference for Detecting Unauthorized Data Use in Code Pre-trained Language Models}}.
\newblock \bibinfo{journal}{\emph{arXiv preprint arXiv:2312.07200}} (\bibinfo{year}{2023}).
\newblock


\end{thebibliography}

\end{document}